\def\etal{{\it et al.\thinspace}}
\def\eg{{\it e.g.\ }} 
\def\ie{{\it i.e.\ }}

\documentclass[10pt,manuscript]{aastex}
\usepackage{graphics}

\shorttitle{CMB Weak lensing at high $\ell$}
\shortauthors{Fullana et al.}

\begin{document}


\title{Estimating small angular scale CMB anisotropy with
high resolution N-body simulations: weak lensing}

\author{M.J. Fullana\altaffilmark{1}, J.V. Arnau\altaffilmark{2},
R.J. Thacker\altaffilmark{3}, H.M.P. Couchman\altaffilmark{4}, and 
D. S\'aez\altaffilmark{5} 
}
\email{diego.saez@uv.es}

\altaffiltext{1}{Institut de Matem\`atica Multidisciplin\`aria, 
Universitat Polit\`ecnica de Val\`encia, 46022 Val\`encia, Spain}
\altaffiltext{2}{Departamento de Matem\'atica Aplicada, Universidad de Valencia, 
46100 Burjassot, Valencia, Spain}
\altaffiltext{3}{Department of Astronomy and Physics, Saint Mary's University,
Halifax, Nova Scotia, B3H 3C3 Canada}
\altaffiltext{4}{Department of Physics and Astronomy, McMaster University,
1280 Main St. West, Hamilton, Ontario, L8S 4M1, Canada}
\altaffiltext{5}{Departamento de Astronom\'{\i}a y Astrof\'{\i}sica, 
Universidad de Valencia, 46100 Burjassot, Valencia, Spain}

\begin{abstract}

We estimate the impact of weak lensing by strongly nonlinear
cosmological structures on the cosmic microwave 
background. Accurate calculation of large
$\ell $ multipoles requires N-body 
simulations and ray-tracing schemes with both high spatial and
temporal resolution. To this end we 
have developed a new code that combines a 
gravitational Adaptive Particle-Particle, Particle-Mesh (AP3M) solver 
with a weak lensing evaluation routine. The lensing deviations are 
evaluated while structure evolves during the simulation so that all 
evolution steps--rather than just a few outputs--are used in the lensing 
computations. The new code also includes a ray-tracing 
procedure that avoids periodicity effects in a universe that is
modeled as a 3-D torus in the standard way. Results from our new simulations are 
compared with previous ones based on Particle-Mesh simulations. 
We also systematically investigate the impact of box volume, resolution,
and ray-tracing directions on the variance of the computed power 
spectra. We find that a box size of $512 h^{-1}$ Mpc is sufficient
to provide a robust estimate of the
weak lensing angular power spectrum in the $\ell $-interval
(2,000--7,000). For a reaslistic cosmological model the power $[\ell(\ell+1)C_{\ell}/2\pi]^{1/2}$ takes on 
values of a few $\mu K$ in this interval, which suggests that a 
future detection is feasible and may explain the
excess power at
high $\ell$ in the 
BIMA and CBI observations.

\end{abstract}


\section{Introduction}
\label{sec1}

Photons from the last scattering surface of the cosmic microwave 
background (CMB) detected at redshift zero, are inevitably lensed by 
cosmological structures.
This lensing produces a number of modifications to the CMB angular power 
spectrum, deviations from Gaussianity, and $B$-polarization. While these effects
have been extensively studied, and a review of their impact can be found 
in \citet{lew06}, we are 
here concerned with the lensing due to strongly nonlinear
objects such as galaxy clusters and smaller-scale structures. In particular, we focus 
our 
attention on the estimation of the small angular scale (high $\ell$) 
correlations induced by this lensing. This marks a somewhat 
different direction to the extensive recent efforts to extend 
N-body simulation 
techniques to all-sky modelling (\eg 
\citealt{f08,db08,car08,t09,car09}).

N-body simulations are necessary to estimate some CMB anisotropies. This 
is the case for the gravitational anisotropies (weak lensing and 
Rees-Sciama effects) produced by strongly nonlinear cosmological 
structures. While simulations with sufficient resolution to model the 
formation of galaxies within an entire Hubble volume are currently out 
of reach, researchers are getting close to this goal 
\citep{k08,t09}. 
However, until such simulations are possible in order to estimate 
gravitational anisotropies in the CMB it is necessary to create 
 an artificial 
periodic universe where copies of the simulation box at various 
redshifts are replicated back to a given redshift. While numerous 
alternatives exist for how 
this is achieved all the methods propagate CMB photons along
 appropriate directions within these model universes. 
When 
conducting this ray-tracing, box sizes, spatial scales and ray 
directions must be chosen in such a way that the periodicity of the 
artificial universe has no appreciable effects on the resulting 
anisotropy. The ray-tracing procedure used here achieves these goals and 
was described and applied in earlier papers, where it was also compared 
with previous methods \citep{jai00,whi00}. Two of these earlier papers 
dealt with weak lensing \citep{cer04,ant05}, and two others with the 
Rees-Sciama effect \citep{sae06,puc06}. The simulations discussed in 
these papers were conducted with the Particle-mesh (PM, \eg 
\citealt{hock88}) N-body simulation algorithm. For the large 
simulation boxes required in our study (see below) the spatial 
resolution of PM simulations is moderate, at best. For this reason it 
was suggested in these papers that better N-body simulations might be 
necessary to estimate good CMB angular power spectra for large multipole 
indexes ($\ell $ values). This suggestion is confirmed in the present 
study, where AP3M (Adaptive Particle-Particle, Particle-Mesh; 
\citealp{co91}) simulations with high resolution are performed to build 
up an appropriate periodic universe, in which the CMB photons are then 
moved according to the aforementioned ray-tracing procedure. A CMB 
angular power spectrum has been estimated from $\ell \sim 200$ to $\ell 
\sim 10000$, and the accuracy of this spectrum in various 
$\ell$-intervals is discussed.

Motivated by a desire to match a number of cosmological constraints, 
including the five-year WMAP 
(Wilkinson microwave anisotropy probe) observations, 
high-redshift supernovae measurements and 
galaxy surveys, (\eg  
Spergel \etal 2007; Vianna \& Liddle 1996; Riess \etal 1998; 
Perlmutter \etal 1999) we consider a cosmological model in which the 
universe is flat and the initial fluctuation spectrum has an 
inflationary origin.
It is assumed that only scalar perturbations 
are present and that weak lensing from gravitational waves is
negligible. 
The resulting distribution of perturbations is then Gaussian 
and the perturbations themselves are purely adiabatic.
Post inflation, the power spectrum of the perturbations is almost 
exactly of the Harrison-Zel'dovich \citep{harr70,zel70} form. This power 
spectrum is then modified during  
the radiation-dominated phase, as acoustic oscillations in the 
photon-baryon fluid damp growth, an effect accounted for by our use of a
CMBFAST calculation of the transfer function. The following 
cosmological 
parameters are used for consistency with the standard model 
\citep{kom08,hin08}: 
(1) a reduced Hubble constant                                                  
$h=10^{-2}H_{0}=0.7$ (where $H_{0}$ is the                                         
Hubble constant in units of km s${}^{-1}$ Mpc${}^{-1}$); (2) 
density parameters $\Omega_{b} = 0.046$, $\Omega_{d} =0.233$, and  
$\Omega_{\Lambda} = 0.721$ for the baryonic, dark matter, and
vacuum energy, respectively (the matter density parameter is then
$\Omega_{m} = \Omega_{b} + \Omega_{d} =0.279$); (3)
an optical depth  $\tau = 0.084$ which characterizes the reionization; 
and (4) a parameter $\sigma_{8} = 0.817 $ normalizing 
the power spectrum of the energy perturbations.  

The layout of this paper is as follows: in Section \ref{maps} we outline 
salient issues in our map-making technique. We follow this in Section 
\ref{nbody} with an outline of N-body methods and our ray-tracing 
technique. Results from our new simulations are presented in Section 
\ref{results} along with a critical comparison to earlier work. We 
then compare our results to current observations and 
conclude with a brief discussion.

\section{Map Construction}\label{maps}

We begin by describing the procedure for constructing lensed maps
of the CMB from small unlensed maps at the last scattering surface. Note that in
this section and those following, we choose units such that 
$c=8\pi G =1$, where $c$ is the
speed of light and $G$ the gravitational constant.
For a quantity $A$, $A_{e}$ and $A_{0}$ denote
the value of $A$ at, respectively, the time of emission  
(last scattering surface) and the present. The scale factor is $a(t)$, where $t$ is the
cosmological time, and its present
value, $a_{0}$, is assumed to be unity, which is always possible 
in flat universes. The unit vector $\vec{n}$ defines the 
observation direction (line of sight).

Small, unlensed maps of CMB temperature contrasts ($\Delta = \delta T/T$) must be 
constructed to be subsequently deformed by lensing. These maps have been obtained 
with a method based                          
on the fast Fourier transform \citep{sae96} and lead                          
to small squared-Gaussian maps of the contrast $\Delta $. These maps are uniformly pixelised. The code 
designed to build up the unlensed maps (map making procedure), requires 
the CMB angular power spectrum, which has been obtained by running 
the CMBFAST code \citep{sel96} for the model described above.
In order to deform the unlensed maps, the lens deviations 
corresponding to a set of directions, covering an 
appropriate region of the sky, must be calculated. These 
deviations are the quantities \citep{selj96}:
\begin{equation}
\vec{\delta} = -2 \int_{\lambda_{e}}^{\lambda_{0}} 
W(\lambda) \vec {\nabla}_{\bot } \phi \ d \lambda \ ,
\label{devi}
\end{equation}
where $\vec {\nabla}_{\bot } \phi = - \vec{n} \wedge \vec{n}
\wedge \vec{\nabla} \phi$
is the transverse gradient of the peculiar gravitational potential $\phi$, and
$W(\lambda) = (\lambda_{e} - \lambda)/ \lambda_{e} $.
The variable $\lambda$ is 
\begin{equation} 
\lambda (a) = H_{0}^{-1} \int_{a}^{1} \frac {db} {(\Omega_{m0}b+
\Omega_{\Lambda} b^{4})^{1/2}} \ .
\end{equation}                

In the case of weak lensing, the integral
in Eq. (\ref{devi}) is usually integrated along straight paths, ignoring 
the small deflections associated with the lensing effect. This 
approximation, commonly known as the Born approximation, is equivalent 
to keeping first order terms in a positional expansion of the transverse 
potential 
as a function of the normal ray and its lensing offset. While this 
approximation is known to be comparatively inaccurate at small $\ell$ 
(\eg
\citealt{vw01}), at $1000\lesssim\ell \lesssim 10000$ detailed 
calculations 
using higher order perturbation theory suggest 
that 
corrections to the first order assumption are approximately two orders 
of magnitude lower than the first order term \citep{sc06}. Recent
N-body simulation work examining the validity of the Born approximation 
(\citealt{h09}) has shown corrections only begin to become 
significant at the 5\% level for $\ell>20000$. Since 
we are 
interested in calculating $1000<\ell<10000$ values we cautiously 
accept the errors inherent in the first order approach.  
 
Once the deviations have been calculated, they can be easily used to get 
the lensed maps from the unlensed ones. This is achieved using the 
relation 
\begin{equation}
\Delta_{_{L}} (\vec {n}) =
\Delta_{_{U}} ( \vec {n} + \vec {\delta} ) \ , 
\label{basmap}
\end{equation}  
where $\Delta_{_{L}} $ and $\Delta_{_{U}}$ are the temperature contrasts of
the lensed and unlensed maps, respectively.

Given the unlensed map $\Delta_{_{U}} $, and the 
map $\Delta_{_{L}} $ obtained from it after 
deformation by lensing (the lensed map), the chosen power spectrum estimator 
can be used to get the quantities 
$C_{\ell}(U) $ and $C_{\ell}(L) $, whose 
differences $C_{\ell}(LU) = C_{\ell}(L) - C_{\ell}(U)$
can be considered as an appropriate measure of the weak lensing effect on the CMB.
Moreover, a map of deformations $\Delta_{_{D}} = \Delta_{_{L}} -  \Delta_{_{U}}$
can be obtained and these can be analyzed to get another angular power spectrum 
$C_{\ell}(D)$. Since the maps $L$ and $U$ are not statistically 
independent,
the spectra $C_{\ell}(D)$ and $C_{\ell}(LU)$ appear to be very different
(see \citet{ant05} for details). 
Eight hundred unlensed maps are lensed (deformed) by 
using the same $\vec{\delta}$ field and the average 
$C_{\ell}(D)$ and $C_{\ell}(LU)$ spectra are calculated and analyzed.
Both spectra are very distinct measures of the weak lensing effect 
under consideration.
Many times, only the customary oscillating $C_{\ell}(LU)$ spectrum  
is shown; however, in a few appropriate cases, 
the $C_{\ell}(D)$ spectra are displayed.
Our power spectrum estimator 
was described in detail in \citet{arn02} and \citet{bur03}.
Results obtained with this estimator were compared with those 
of the code ANAFAST of the HEALPix \citep{gor99} package 
in \citet{arn02} and also in \citet{puc06}. 
These comparisons showed that our estimator is a very good one 
in the case of regularly pixelised squared maps as analyzed
here.

\section{N-body simulations and ray-tracing procedure}  
\label{nbody}
\subsection{N-body simulation technique}

The primary simulations presented in this work were run using 
 a parallel OpenMP-based implementation of the ``HYDRA'' code 
\citep{tc06}. This code uses the AP3M algorithm to calculate 
gravitational forces within a simulation containing $N_p$ particles. In 
the AP3M algorithm a 
cubic ``base" mesh 
of size $N_c$ cells per side is supplemented by a series of 
refined-mesh P3M calculations to provide sub-mesh resolution. 
Gravitational 
softening is 
implemented using the S2 softening
kernel
\citep{hock88} which is remarkably similar in shape to the cubic
spline softening kernel used in many treecodes (\eg \citealp{hk89}).
The S2 softening used in the kernel is  
$2.34\times S_{p}$ where $S_{p}$ is an equivalent Plummer softening length 
which we 
quote throughout the paper to enable a simple comparison to other work. The
softening length is held constant in physical coordinates subject to the
resolution not falling below 0.6 of the interparticle spacing at high
redshift. This technique is widely applied (\eg \citealp{mill}) and is 
a compromise
between assuring that the potential energy of clusters does not evolve
significantly at low redshift, while still ensuring structures and
linear perturbations at
high redshift are followed with reasonable accuracy.

\begin{deluxetable}{lccccccccccccc}
\tabletypesize{\tiny}
\tablecaption{Lensing Simulations\label{tab2}}
\tablewidth{0pt}
\tablehead{
\colhead{Name} & \colhead{Algorithm} & \colhead{$L_{box}$} & 
\colhead{Mode set} & 
\colhead{$N_{p}$}& $M_{p}$ &
\colhead{$N_c$}& \colhead{$S_{p}$}& \colhead{}& \colhead{$N_{dir}$}& 
\colhead{$z_{in}$ }& 
\colhead{$\Delta_{ps}$ }& $\Delta_{ang}$ & \colhead{Direction} \\
\colhead{}& \colhead{} & \colhead{$/h^{-1} Mpc$} & 
\colhead{}& \colhead{}& \colhead{ $10^{10} M_\odot$} &
\colhead{}& \colhead{$/h^{-1} kpc$}& \colhead{}& \colhead{}& 
\colhead{}& \colhead{$/h^{-1} kpc$}& \colhead{'}& \colhead{}
}
\startdata
RLS\_AA & AP3M & $512$ & A & $512^3$ & 11 & 1024 & 12 &${}^{}$ & 512 & 
6 &
25 & 0.59 & D512A \\
RLS\_AB & AP3M & $512$ & A & $512^3$ & 11 &1024 & 12 &${}^{}$ & 512 & 6 
&
25 & 0.59 & D512B \\
RLS\_BA & AP3M & $512$ & B & $512^3$ & 11 &1024 & 12 &${}^{}$ & 512 & 6 
&
25 & 0.59 & D512A \\
RLS\_CA & AP3M & $512$ & C & $512^3$ & 11 &1024 & 12 &${}^{}$ & 512 & 6 
&
25 & 0.59 & D512A \\
LS\_LDA & AP3M & $1024$ & D & $512^3$ & 88 &1024 & 24 &${}^{}$ & 1024 & 
6 &
50 & 0.59 & D1024A \\
LS\_MAB & AP3M & $512$ & A & $256^3$ & 88 & 512 & 24 &${}^{}$ & 512 & 6 
&
50 & 0.59 & D512B \\
LS\_$\Delta$S1AA & AP3M & $512$ & A & $512^3$ & 11 &1024 & 24 &${}^{}$ & 
512 
& 6 &
50 & 0.59 & D512A \\
LS\_$\Delta$S2AA & AP3M & $512$ & A & $512^3$ & 11 &1024 & 36 &${}^{}$ & 
512 
& 6 & 75 & 0.59 & D512A \\
LS\_$\Delta$1AA & AP3M & $512$ & A & $512^3$ & 11 &1024 & 12 &${}^{}$ & 
512 
& 6 & 12 & 0.59 & D512A \\
LS\_$\Delta$1AB & AP3M & $512$ & A & $512^3$ & 11 &1024 & 12 &${}^{}$ & 
512 & 6 & 12 &0.59 & D512B \\
LS\_$\Delta$1BA & AP3M & $512$ & B & $512^3$ & 11 &1024 & 12 &${}^{}$ & 
512
& 6 & 12 & 0.59 & D512A \\
LS\_$\Delta$1BB & AP3M & $512$ & B & $512^3$ & 11 &1024 & 12 &${}^{}$ &
512 & 6 & 12 &0.59 & D512B \\
LS\_$\Delta$2AA & AP3M & $512$ & A & $512^3$ & 11 &1024 & 12 &${}^{}$ & 
512
& 6 & 40 & 0.59 & D512A \\
LS\_$\Delta$2AB & AP3M & $512$ & A & $512^3$ & 11 &1024 & 12 &${}^{}$ &
512 & 6 & 40 & 0.59 & D512B \\
LS\_$\Delta$2BA & AP3M & $512$ & B & $512^3$ & 11 &1024 & 12 &${}^{}$ & 
512
& 6 & 40 & 0.59 & D512A \\
LS\_$\Delta$2BB & AP3M & $512$ & B & $512^3$ & 11 &1024 & 12 &${}^{}$ &
512 & 6 & 40 & 0.59 & D512B \\
LS\_HEA  & AP3M & $256$ & E & $512^3$ & 1.4 & 1024  & 6 &${}^{}$ & 512 & 
6 
&
15 & 0.29 & D256A \\
LS\_MGA & AP3M & $512$ & G & $256^3$ & 88 & 512 & 24 &${}^{}$ & 512 &
6 & 60 & 0.59 & D512A \\
LS\_LHA & AP3M & $512$ & H & $128^3$ & 704 & 256 & 48 &${}^{}$ & 512 &
6 & 120 & 0.59 & D512A \\
PMLS & PM  & $256$ & F & $512^3$ & 1.4 & 512 & 1000 &${}^{}$ & 256 & 6 
& 500 & 0.59 & D256A \\
\enddata
\tablecomments{$L$ist of the parameters used in the lensing 
simulations. 
Columns three to eight give information associated with the N-body 
simulations, namely the box size, $L_{box}$, the set of initial modes 
(determined by the box size and a single random seed), the 
number of particles used, $N_{p}$, the mass of the individual 
particles, $M_{p}$, the number of Fourier cells along 
each side of the simulation, $N_c$, and the Plummer softening length 
used, 
$S_{p}$. All N-body simulations were started at a redshift of $z=50$. 
The 
following five columns give parameters associated with the lensing 
calculation, namely the number of rays along the edge of the map, 
$N_{dir}$, the initial redshift at which ray tracing begins, $z_{in}$, 
the 
distance between evaluations on the geodesic $\Delta_{ps}$, 
the angular resolution of the CMB lensed maps $\Delta_{ang}$, and the 
preferred direction used (see Table 2).}
\clearpage
\end{deluxetable}

Initial conditions were calculated using the standard Zel'dovich 
approximation technique \citep{efsta85}, and all simulations were 
started at a redshift of $z=50$, which is sufficiently early to place 
modes in the linear regime. To account for the impact of varying box 
sizes, $L_{box}$, we considered simulations of size $256 h^{-1}$ Mpc, 
$512 h^{-1}$ Mpc, and $1024 h^{-1} $ Mpc and a full list of 
N-body simulation and ray-tracing
parameters is given in Table~\ref{tab2}. At $z=6$ (the 
beginning of our ray-tracing epoch in the simulation), the box sizes of 
$256 h^{-1}$ Mpc, $512 h^{-1}$ Mpc, and $1024 h^{-1}$ Mpc correspond to 
square sky patches with angular sizes, $\Phi_{map}$, of $2.48^{\circ}$, 
$4.96^{\circ}$, and $9.92^{\circ}$, respectively. These are then the 
sizes of our constructed CMB maps for each simulation.

\subsection{Ray-tracing technique}
\subsubsection{Lensing regimes}
As in our previous work \citep{ant05} we divide the total lensing effect 
into three parts:
\begin{itemize}
\item AWL (A weak lensing), namely the effect due to scales $k > 2\pi / 
L_{max} $ (where $L_{max} = 42 h^{-1}
$ Mpc) at redshifts $z < 6 $. This signal is dominated by strongly 
nonlinear scales
\item BWL, the lensing signal due to scales $k < 2\pi / L_{max} $ which 
corresponds to modes that are always in the linear regime down to $z=0$ 
\item CWL, the lensing signal due to scales  $k \geq 2\pi / L_{max} $ 
but at redshifs $z>6$
\end{itemize}

The goal of this paper is to calculate the AWL signal, using ray-tracing 
through N-body simulations, to $\ell$ values in the range $1000-10000$. 
Since the modes associated with the BWL component are specifically 
chosen to
correspond to scales that are linear down to $z=0$, which is set by
wavenumbers $k\lesssim 0.15 h$ Mpc$^{-1}$ \citep{sm03}, the BWL 
component 
can be calculated with the linear approach implemented in CMBFAST. The CWL 
component involves modes in the mildly nonlinear regime which can, 
nonetheless, be evolved via approximation schemes 
\citep{zel70,san89,mou91}. Standard semi-analytical
methods designed to study weak lensing in the nonlinear regime 
should also apply in this case, hence, the CWL effect can be 
calculated without resorting to N-body techniques. This calculation can be 
performed using the nonlinear version of CMBFAST which is based on 
semi-analytical approaches \citep{lew06}.
For these reasons we begin ray tracing within the
simulation at $z=6$ and consider nonlinear spatial scales having
wavenumbers $k \geq 2\pi / L_{max} $.
Other methods utilizing simulations have also included the 
contribution to lensing from all scales above a certain redshift using 
special techniques 
\citep{db08,car08}. 

\begin{figure}[h!]
\epsscale{.70}
\plotone{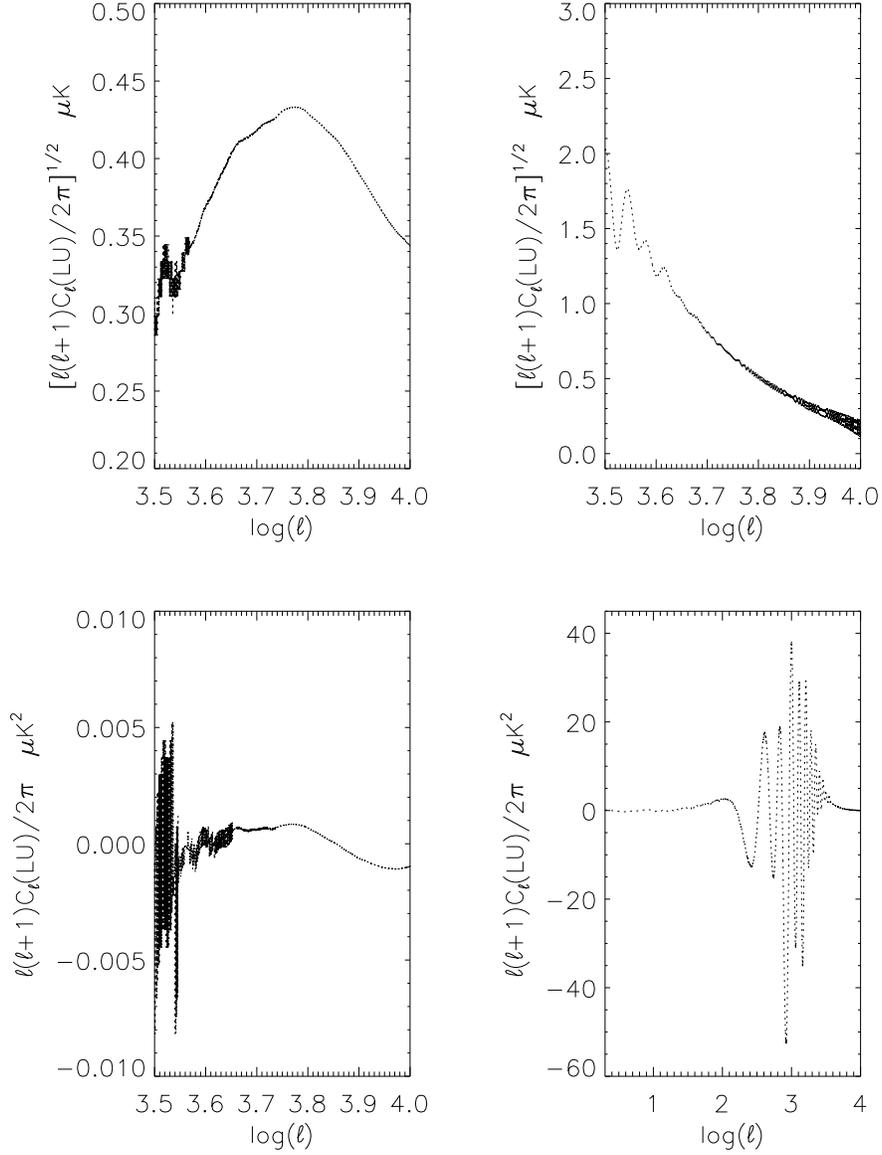}
\caption{Results from CMBFAST calculations.
Top-left, top-right and bottom
left panels exhibits the LU angular power spectrum for
$\ell > 3200 $ in the cases AWL, BWL, and CWL, respectively.
Bottom-right panel is the LU spectrum corresponding to the
BWL effect for $\ell < 10000 $. This effect dominates against AWL and
CWL for
$\ell < 3000$.
\label{fig0}}
\end{figure}

In Figure~\ref{fig0} we show calculations of the LU angular power 
spectra for the AWL, BWL and CWL lensing 
regimes 
for $3200<\ell<10000$ using the nonlinear lensing 
implementation in CMBFAST. Note that for $\ell<3500$ the AWL and CWL 
effects are too small to be calculated with CMBFAST due to the fact 
that these spectra are obtained as differences between other spectra 
provided by the code. Both the AWL and CWL effects, as calculated by 
CMBFAST, 
are clearly small with the CWL effect being virtually negligible in the 
$3200<\ell<10000$ range. As anticipated, the BWL effect is dominant. 
The calculation of the AWL effect provides an initial estimate 
of the signal we wish to calculate with the simulation. We have verified
that, for $\ell<10000$, the BWL effects  
calculated by using both the linear and nonlinear methods 
implemented in CMBFAST are almost identical, which proves 
that the BWL component is linear, as expected.

\subsubsection{Calculation of the lens deviation integral}
 An estimate of the lens deviation 
integral, given in Eq.~(\ref{devi}), is computed via a numerical 
integration performed along the background null geodesics from the 
comoving distance $D_{max} \simeq 5900 h^{-1} \ Mpc$ corresponding to 
$z=6$ (in the model under consideration) to the observer position.  The 
gradient of the peculiar potential $\vec{\nabla} \phi$ used in the 
integrand of Eq.~(\ref{devi}) is obtained from the simulation but is not 
exactly the same as that used within the N-body calculation. Instead, 
the gradient is found by subtracting the part of the full N-body 
potential produced by linear spatial scales larger than $42 h^{-1} \ 
Mpc$.

Specifically, our algorithm for determining the potential gradient, 
which is proportional to the force, is as follows:
\begin{enumerate}
\item Decide upon the direction of the normal rays representing the 
geodesics (see Section~\ref{pref})
\item Assuming the Born approximation and using the photon step distance 
$\Delta_{ps}$ (see Section~\ref{pref}) determine all 
the evaluation 
positions and times on the geodesics within the simulation volume from 
$z=6$ down 
to the final redshift
\item Associate test particles with each of these positions and times
\item At each time-step of the N-body simulation (while it is 
running) determine which test particles require force evaluations
\item At each test particle position evaluate the force on the test 
particle using the long-range FFT component and short-range PP 
correction as in the HYDRA algorithm
\item During the FFT convolution for the test particles eliminate 
contributions from scales 
larger 
than  $42 h^{-1}$  Mpc by removing the signal from wavenumbers 
satisfying  $k\lesssim 0.15 h$ Mpc${}^{-1}$
\item If the evaluation time for a point on the geodesic lies between 
two 
time-steps calculate a linear interpolation of the two forces from the 
time-steps that straddle the correct time
\item Resolve the force into its transverse component and hence recover 
the transverse component of the potential gradient
\end{enumerate}

Once all the potential gradients are evaluated we can calculate the 
lensing deviation integrals. We emphasize that our algorithm ensures the 
potential gradient along 
normal rays is  
calculated very accurately and corresponds exactly (modulo the removal 
of power from scales greater than $42 h^{-1}$ Mpc) to that in the 
simulation. We do not resort to smoothing onto grids or 
the creation of lensing planes ``on the fly''. The main drawback of our 
method is that since lensing is now an integral part of the 
N-body simulation, 
a different preferred direction 
requires a new N-body
simulation.

\subsubsection{Photon propagation paths}
\label{pref}

To calculate the photon propagation, the CMB photons are moved through 
the simulation volume along specially chosen paths to avoid repeatedly 
sampling the same structures. This approach uses all steps from the 
simulation and by using the periodicity of the box volume there are no 
discontinuities in the matter field anywhere. In principle this approach 
is similar to tiling methods that are used elsewhere (\eg 
\citealt{whi00,h01,s09}). These methods are based on independent
PM simulations with decreasing sizes that telescope in 
resolution along the line of sight. In our ``tiling''
the matter field is constantly being    
updated by the AP3M code 
as the 
time-step changes and, consequently, it is 
not limited by the box size. 
We 
also take care to ensure paths are taken which avoid, as much as 
possible, periodicity effects. This is notably different from other 
approaches using random translations and orientations (\eg 
\citealt{car08}) of the simulation volume which, unavoidably, have 
discontinuities at adjoining radial shells. While the signal from such 
discontinuities is likely small, our method has the advantage 
of avoiding it completely.

Choosing the directions for the ray propagation is not entirely trivial. 
For the directions parallel to the 
box edges, periodicity effects are clearly very strong. Photons moving 
along these directions
would pass close to the same structures in successive boxes and,
consequently, the lensing effect of the same structure would be 
included a number of times (one per crossed box). This repetition
would lead to a false magnification of lens deviations.
However, as has been emphasized in \citet{ant05} and \citet{sae06},
periodicity effects are negligible
along certain directions, hereafter called 
{\em preferred directions}. 

In order to define preferred directions, it can be assumed that: (i) the 
$x$, $y$, and $z$ axes are parallel to the box edges; (ii) the angles 
$\theta $ and $\varphi $ are spherical coordinates defined with respect 
to these axes; (iii) photons moving along the direction $\vec{n}$, cross 
the $(y,z)$ face of a box at point $P$, and the next $(y,z)$ face at 
point $Q$; and (iv) if the segment PQ is projected onto the $(y,z)$ 
plane, the length of the resulting projection is $\zeta_{_{PQ}}$. Then, 
if the condition $\zeta_{_{PQ}} > L_{max} $ is satisfied, the direction 
$\vec{n} $ is assumed to be a preferred one. Taking into account this 
definition of preferred directions and the cutoff performed at the scale 
$L_{max} $, it is clear that photons moving along preferred directions 
enter successive boxes through independent uncorrelated regions. 
Moreover, for these directions and the box sizes of interest, it can be 
easily verified that the CMB photons can travel from $z=6$ to $z=0$ 
($\sim 5900 h^{-1} \ Mpc$) through different uncorrelated regions 
(without repetitions) located in successive simulation boxes. 
Periodicity effects can, therefore, be assumed to be negligible. The 
total number of crossed boxes is denoted by $N_{cr}$.

Some preferred directions used in this paper are defined in Table \ref{tab1}, 
where the first column gives the names of all these directions. 
Each of them has been used in universes
covered by simulation boxes with a given size, as detailed 
in the second column. The third and fourth columns show, respectively, the 
angles $\theta$ and $\varphi$ (spherical 
coordinates) defining the corresponding direction. 
The distance     
$\zeta_{_{PQ}}$ defined in the previous paragraph is given 
in the fifth column and, finally, in the last column the 
number of boxes crossed by the CMB photons from redshift 
$z=6 $ to the present is given. From the values of Table~\ref{tab1}, it is 
clear that 
the distance
$\zeta_{_{PQ}}$ is much greater than 
$L_{max} = 42 h^{-1} \ Mpc$ in all four cases, which ensures that 
other directions close enough to the listed ones 
are also preferred directions. 

\begin{deluxetable}{cccccc}
\tablecaption{Preferred directions for ray tracing.\label{tab1}}
\tablewidth{0pt}
\tablehead{
\colhead{Direction} & \colhead{$L_{box}$} & \colhead{$\theta$} & \colhead{$\varphi$} &
\colhead{$\zeta_{_{PQ}}$} & \colhead{$N_{cr}$}
}
\startdata
D256A  & 256 & 76.76 & 11.31 & 79.98 & 22.00 \\
D512A  & 512 & 76.76 & 11.31 & 159.95 & 11.00 \\
D512B  & 512 & 68.43 & 18.435 & 273.20 & 10.17 \\
D1024A & 1024 & 59.19 & 26.57  & 853.33 & 4.43 \\
\enddata
\tablecomments{Distances $L_{box}$ and $\zeta_{_{PQ}}$ are given in units of
$h^{-1} \ Mpc$, and the spherical coordinates $\theta $ and $\varphi$ in degrees}
\end{deluxetable}
\clearpage

We proceed
as follows: any direction of Table~\ref{tab1} is assumed to point 
toward the center of a squared map, 
whose angular size corresponds to the box size appearing in 
the second column (as noted above). 
Then, these squared maps are uniformly 
pixelised by choosing a certain number of pixels, $N_{pix}$, 
per edge. The angular resolution is then $\Delta_{ang} = \Phi_{map} /
N_{pix}$. The directions of all the pixels are preferred ones and,
consequently, lens deviations can be calculated 
for each pixel---with no significant periodic effects---across the full map.

The parameters involved in the ray-tracing procedure are thus summarized:
a number of directions, $N_{dir}$, per edge of the squared CMB map 
(one per pixel, $N_{dir}=N_{pix}$); 
an initial redshift, $z_{in}$, for the calculation of lens deviations; 
a step, $\Delta_{ps} $, to perform the integral in Eq.(\ref{devi}) 
(hereafter called the photon step); and the angles $\theta$ and $\varphi$ 
defining the preferred direction.

Hereafter, a lensing simulation (LS) is the 
calculation of the $\vec{\delta}$-deviations 
along the pixel directions, plus the construction of the 
$\Delta_{_{U}}$, $\Delta_{_{L}}$, $\Delta_{_{D}}$ maps and 
the estimation of $C_{\ell}(D)$ and $C_{\ell}(LU)$
(angular power spectra). An LS is characterized by the parameters
and initial conditions required by the N-body simulation  
together with the parameters of the ray-tracing 
procedure. 
The lensing simulations (LSs) obtained from the parameters: 
$L_{box} = 512 h^{-1} \ Mpc $, $N_{p} = 512^{3}$, $N_{c} = 1024$, 
$S_{p} = 12 h^{-1} \ kpc$, 
$N_{dir} = 512$, 
$z_{in}=6$ (as in all the simulations used in the paper), and
$\Delta_{ps} = 25  h^{-1} \ kpc$, are hereafter
called reference lensing simulations (RLSs). 
The angular resolution of these simulations is $\Delta_{ang} \simeq 0.59^{\prime}$
($\ell \simeq 18,600$). There
are an infinite number of possible realizations of this type
of LSs corresponding to different initial
conditions for the N-body simulation as well as to distinct preferred directions. We 
also consider the 
effect on the LSs of changing parameters to ensure that the calculation of 
the 
power spectra $C_{\ell}(D)$ and $C_{\ell}(LU)$ is as robust and accurate as possible.

\section{Results}
\label{results}
\subsection{Effect of changing the final ray-tracing redshift}                 
\begin{figure}
\epsscale{.80}
\plotone{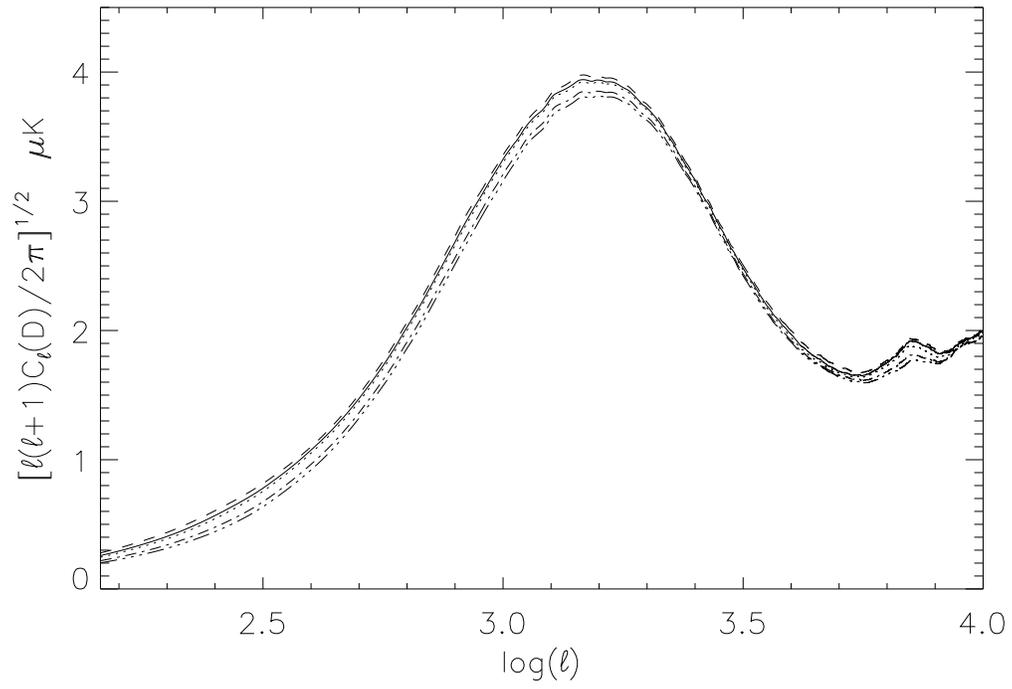}
\caption{$D$ angular power spectra corresponding to 
RLS\_AA at different final redshifts. Curves corresponding to $z_{end}$ 
values of 0.5 (triple-dot-dash), 0.4 (dot-dash),
0.3 (dots), 0.2 (solid), and 0.0 (dashes) are shown.
\label{fig1}}
\end{figure}

To determine the impact of changing the final ray-tracing epoch, the 
RLS\_AA lensing simulation (based on the preferred direction $D512A$, 
see tables \ref{tab1} \& \ref{tab2}), was used to estimate lensing from 
$z_{in}=6$ to final redshifts between $0.5$ and $0$. Results are shown 
in Fig.~\ref{fig1}, where the $D$ angular power spectra corresponding to 
the final redshifts 0.3 (dotted line), 0.2 (solid line), and 0.0 (dashed 
line) are presented. Clearly, these three spectra are very similar. We 
find the same result for the $LU$ spectra (these results are omitted 
since the addition of five oscillating lines would lead to a confusing 
plot and add little information).
Based upon these results we conclude that the signal produced between 
redshifts $0.2$ and $0$ contributes negligibly to the total lensing. 
Given this fact, and that the CPU time required to evolve between 
$z=0.2$ and $z=0$ is comparatively lengthy for our code (due to the 
absence of individual particle time-steps), our calculations are 
performed between redshifts $z_{in}=6$ and $z_{end} = 0.2$. The same 
study has been done for other RLSs corresponding to different 
initial conditions and preferred directions. The conclusions are the 
same in all the cases, namely a negligible effect between $z=0.2$ and 
$z=0$. We note that this conclusion is in close agreement with that of 
\cite{car08} who showed that stopping at $z=0.22$ produced a deficit in 
the low $\ell$ signal but a negligible difference for $\ell>350$.

\subsection{AP3M simulations contrasted to PM}
\label{comparison}
\begin{figure}
\epsscale{.80}
\plotone{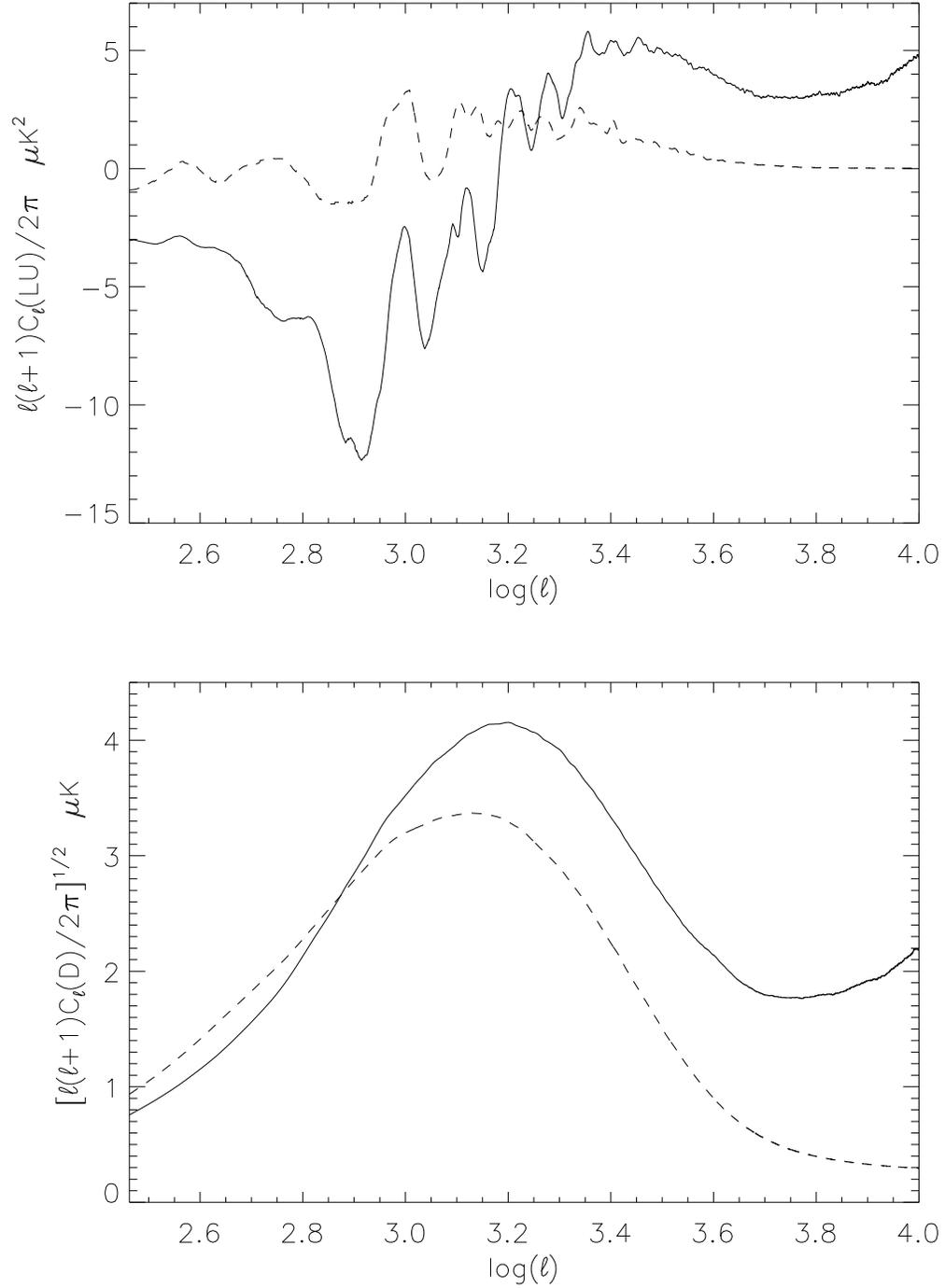}
\caption{Top: $LU$ angular power spectra corresponding to the
RLS\_BA (solid line) LS, and to the PMLS LS defined in the text
(dashed line). 
Bottom: $D$ spectra for the same simulations as in
top panel.
\label{fig2}}
\end{figure}

We next compare LSs constructed from PM and AP3M N-body codes to examine 
the impact of sub-mesh-scale resolution on the lensing signal. In 
Fig.~\ref{fig2} we plot the $C_{\ell}(LU)$ (top panel) and $C_{\ell}(D)$ 
(bottom panel) for RLS\_BA
(solid line) along with the same spectra for PMLS (dashed line),
obtained from a PM code. Details of the PM code can be found
in \cite{qui98}; this code has been used in previous papers (\eg
\citealp{ant05}). We emphasize that the same
 ray-tracing procedure, described in Section~\ref{nbody}, is used in 
both codes.
The parameters of the PMLS (see Table~\ref{tab2}) are as follows:
$L_{box} = 256 h^{-1} \ Mpc $, $N_{p} = 512^{3}$, $N_{c} = 512$ , 
$N_{dir} = 256$, and $\Delta_{ps} = 0.5 h^{-1} \ Mpc$, and the preferred direction 
is D256A (see Table~\ref{tab1}).
The effective resolution $E_{res}$ of the PM code is 
two cells (1 $h^{-1} \ Mpc$ in this case); hence, in the PMLS, 
the photon step $\Delta_{ps}$ must be 
smaller than $1 h^{-1} \ Mpc$ to take advantage of the 
PMLS resolution. It has been verified that                                                              
$\Delta_{ps} = 0.5 h^{-1} $ Mpc (half of $E_{res}$) is a good value for the 
photon step (smaller values lead to very similar results).
In keeping with other N-body work (\eg \citealp{moore98}), the 
true effective resolution of the AP3M simulations 
is estimated to be 
$E_{res} \sim 5S_{p}$, which means that, in our  
RLSs, the photon step is a little smaller than a half of the 
effective resolution. 

Examination of Fig. \ref{fig2} shows that the PMLS traces the peaks, but 
the amplitudes are too small and, moreover, the $C_{\ell}(LU)$ 
quantities quickly tend to zero as $\ell $ increases. However, in the 
AP3M case, the peaks have greater amplitudes and, furthermore, a signal 
of a few micro-Kelvin appears for high $\ell $ values. Similar 
conclusions follow from the $D$ spectra of the bottom panel, which can 
be directly compared with \citet{ant05}, where this kind of spectrum was 
used. These results show that the PMLS underestimates the lensing signal 
we are calculating. Hence, high resolution N-body simulations, 
be they AP3M or computed using some alternative algorithm, are necessary 
to estimate CMB weak lensing from strongly nonlinear structures.

\subsection{Variance of power spectra due to modes in the initial conditions}
\label{modes}
\begin{figure}
\epsscale{.80}
\plotone{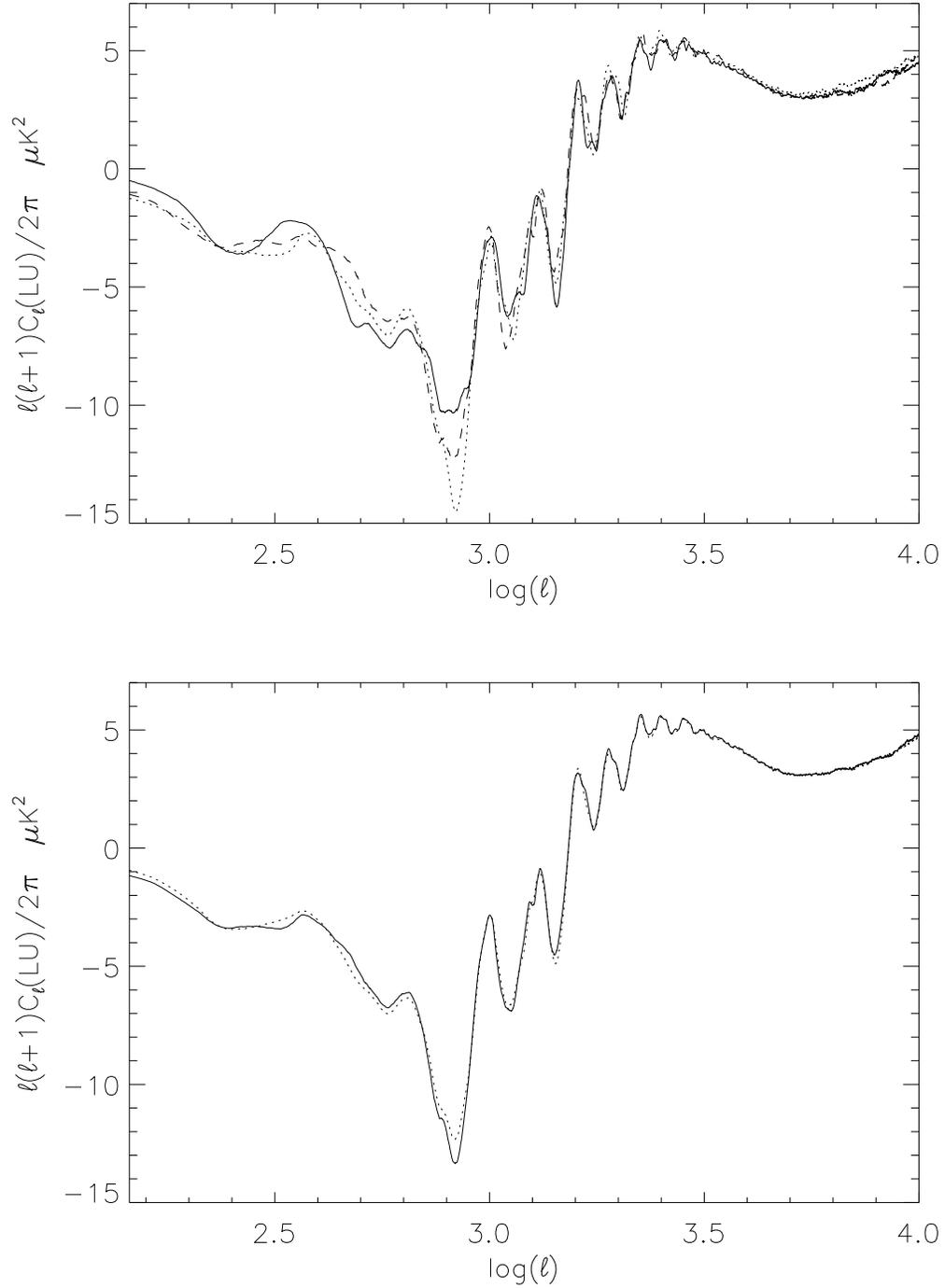}
\caption{Top: Three $LU$ angular power spectra extracted from distinct 
RLSs, namely RLS\_AA (solid), RLS\_BA (dashed), RLS\_CA (dotted). 
Bottom: the dotted line gives the average $LU$ spectrum of the three 
RLSs of the top panel, whereas the solid line corresponds to the 
average of two spectra, namely just RLS\_BA and 
RLS\_CA.
\label{fig3}}
\end{figure}

Having shown that the results from the AP3M simulations are distinctly 
different from those of the PM simulations, we must still account for 
variability of results due to changing the random modes in the initial 
conditions of the simulation. To this end we show, in the top panel of 
Fig.~\ref{fig3}, the $LU$ angular power spectra corresponding to three 
distinct RLSs, namely, RLS\_AA, RLS\_BA \& RLS\_CA, each drawn from a 
different 
realization of the modes but 
using the same preferred direction ($D512A$ in Table~\ref{tab1}). The 
results are clearly qualitatively very similar and the only noticeable 
quantitative differences are below $\ell\sim1000$. This is not entirely 
surprising since the lower the $\ell$ value the larger the sample 
variance \citep{kno95, sco94}. Additionally, the largest deviation 
occurs at the point where the signal is weakest, again an unsurprising 
result.

These results suggest that a good average $LU$ spectrum can be achieved 
from just a few chosen RLSs simulations, perhaps as few as two. We have 
verified this assertion in the bottom panel of Fig.~\ref{fig3}, by 
plotting the average of two of the RLSs as compared to the average of 
all three RLSs. Both average spectra are so similar that only two RLSs 
suffice to get a very good average $LU$ spectrum in the $\ell$-interval 
under consideration. Unsurprisingly, the average signals for $\ell < 
1000 $ are in good agreement, as would be expected if the differences in 
the single realizations were due to sample variance.
 For $\ell > 1000 $, the spectra of the different realizations are 
so similar that
a unique RLS leads to a rather good $LU$ spectrum of the lensing effect
under consideration ($L<L_{max}$ and $z<z_{in}=6$).

We emphasize that the sample variance we have been talking about is 
inherent in the lensing calculation itself, rather than the CMB maps 
themselves. For each RLS, the spectra are calculated from $800$ small 
maps having a size of $4.96$ degrees; which is a necessary step to 
understand small sample variances (small deviations from spectrum to 
spectrum).

\subsection{Variance of power spectra due to different preferred directions}

\begin{figure}
\epsscale{.80}
\plotone{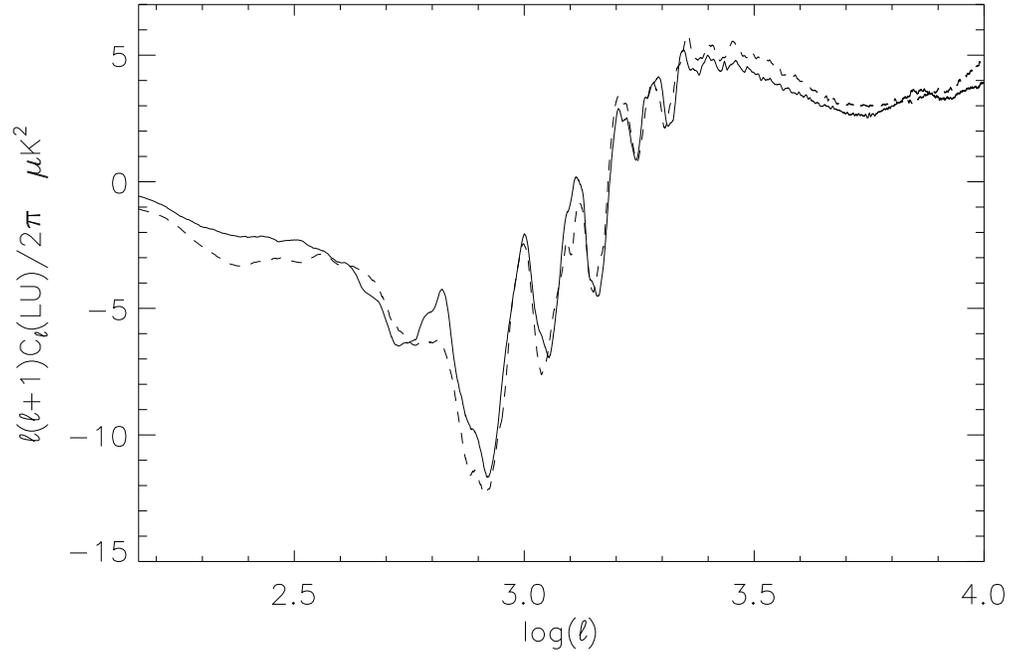}
\caption{$LU$ angular power spectra obtained from RLS\_AA and RLS\_AB
which share the same N-body initial conditions 
but have distinct preferred directions. The dashed (solid) line 
corresponds 
to RLS\_AA (RLS\_AB) and hence the direction D512A (D512B).                   
\label{fig4}}
\end{figure}

In terms of the variance produced by different preferred directions, the 
ergodic nature of the ray-tracing means that we might reasonably expect 
the same level of variance seen for the different N-body realizations 
with differing modes. Nevertheless, choosing a different preferred 
direction for a given RLS means that the photons 
move through different 
regions of the simulation box and
cross box faces in different places 
and a different number of times. So the trajectories are not fully 
equivalent, in turn producing lensing deviations that are subtly 
distinct. To evaluate this difference we plot, in Fig.~\ref{fig4}, two 
$LU$ angular power spectra for the RLSs RLS\_AA and RLS\_AB which use 
the same 
initial modes, but have preferred directions D512A and 
D512B. The similarity of the two 
curves indicates that, as anticipated, results are almost independent of 
the chosen preferred direction, and that our ray-tracing procedure is 
robust.

\subsection{Impact of simulation box size on the power spectra}
Thus far we have considered AP3M simulations with box sizes 
 $L_{box}=512h^{-1}$  Mpc, and hence the $LU$ and $D$ spectra have
 been obtained from small maps of angular 
size
$\Phi_{map}=4.96^{\circ}$. We next examine whether these values are
too small to make robust predictions.

\begin{figure}
\epsscale{.80}
\plotone{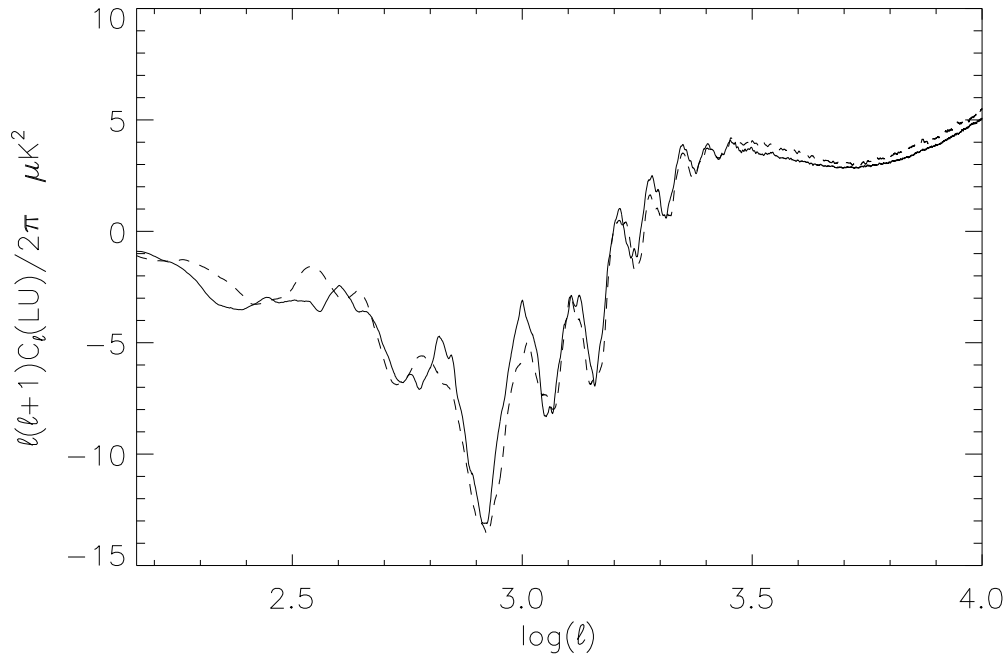}
\caption{LU angular power spectra extracted from two LS
simulations (LS\_LDA, solid line and  LS\_MAB, dashed line) that differ 
in 
box sizes but 
use similar 
effective particle 
and map resolutions. The simulations share the same softening and photon 
steps, namely $S_{p} = 24 h^{-1} \ kpc$ and
$\Delta_{ps} = 60 h^{-1} \ kpc$, but for the LS\_LDA LS 
$L_{box}=1024 h^{-1} \ Mpc$, 
$N_{p} = 512^{3}$ while for the LS\_MAB LS, $L_{box}=512 h^{-1} \ 
Mpc$, 
$N_{p} = 256^{3}$. 
The initial conditions of these LSs are fully independent and different 
preferred directions are used.
\label{fig5}}
\end{figure}

In order to answer this question, we conducted an LS, labeled LS\_LDA in 
Table~\ref{tab2},
with $L_{box}=1024 h^{-1} \ Mpc$ and $N_{dir}=1024$, which leads to 
a lensed map of size $\Phi_{map} = 9.92^{\circ} $ and with the same
angular resolution as in the RLSs. Since simulating a larger volume at 
fixed particle resolution
requires 
a larger 
softening and the corresponding photon step, the values
$S_{p} = 24 h^{-1} \ kpc$, and 
$\Delta_{ps} = 60 h^{-1} \ kpc$ were used. The preferred direction 
was 
D1024A (see Table~\ref{tab1}). To compare to this simulation we also ran 
another LS, denoted LS\_MAB, with parameters set to mimic the LS\_LDA 
simulation but in 1/8th the volume and using 1/8th the number of 
particles. 
The $LU$ spectra obtained for the two simulations are shown in 
Fig.~\ref{fig5}, and are extremely similar. This is despite the fact 
that these two simulations 
have differing initial conditions, different box sizes, and different numbers of
crossed boxes, $N_{cr}$.   
Hence we conclude that a box size of $512h^{-1} \ Mpc $---the same as 
that
of the RLSs---is 
large enough to get very good angular power spectra for the range of
$\ell$ considered. Even smaller
sizes, \eg, $256h^{-1} \ Mpc$, can be used when 
only large enough $\ell$ values are under consideration (see below).

\subsection{Noise in the power spectra}
Although we do not directly calculate the 2d lensing potential or 
convergence we can examine the power spectrum of the angle
$\alpha = |\vec{\delta}| $   
as in \cite{car08}. At our final redshift of 0.2, the angular resolution 
of our RLSs (0.59') corresponds to a comoving separation of 100$h^{-1}$ 
kpc, which is still larger than our nominal N-body resolution of 
5$S_p=60 h^{-1}$ kpc. Since this is the effective resolution in the 
ray-tracing we do not need to worry about the intrinsic resolution of 
our maps falling below that of the N-body simulation. 

\begin{figure}
\epsscale{.30}
\plotone{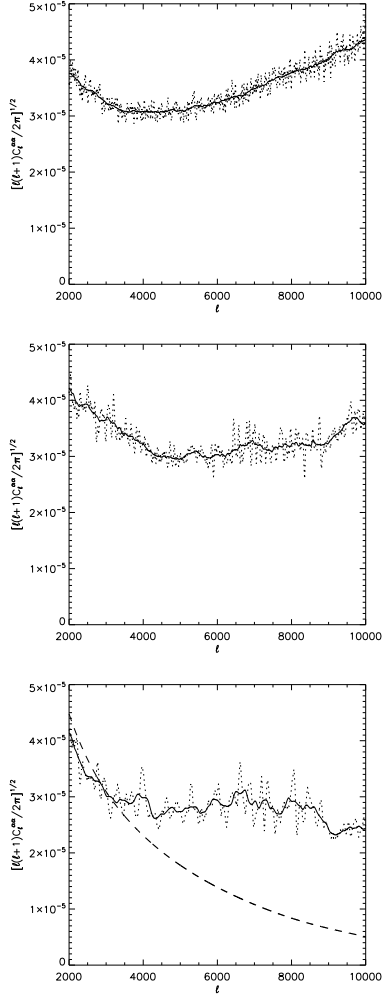}
\caption{Top, medium and bottom panels 
correspond to box sizes of $1024h^{-1}$ Mpc (LS\_LDA), $512h^{-1}$ Mpc 
(RLS\_BA ), 
and $256h^{-1}$ Mpc (LS\_HEA).
Dotted lines correspond to the correlations 
extracted from the 
simulated $\alpha $ maps. Solid lines are averaged spectra in running 
bins of size $\ell\pm100$. 
The mean amplitude and the $rms$ of the relative errors 
(dotted-solid/solid) are
 (-0.00072,0.041),
(-0.00065,0.053), and (-0.00056,0.062) in the top, medium, and bottom 
panels, respectively. The dashed line in the bottom panel corresponds to 
the total lensing signal predicted by CAMB. Note that the solid and 
dotted lines in all panels correspond to the AWL lensing component.
\label{newplot}}
\end{figure}

In Fig.~\ref{newplot} we plot the power spectrum of the lens deviations 
in the interval $2000<\ell<10000$.  We have quantified the noise in this 
signal by considering box sizes of 1024$h^{-1}$ Mpc, 512$h^{-1}$ Mpc, 
256$h^{-1}$ Mpc although our smallest box has an effective angular 
resolution that is twice as small as the other two simulations.
  For comparison to semi-analytic methods we also include the {\em 
total} lensing signal predicted by CAMB in this $\ell$ range. Comparing 
to a running average constructed by binning the average signal in 
$\ell\pm 100$, the RMS deviations relative to the overall signal are 
4.1\%, 5.3\% and 6.2\% for the three box sizes, respectively. Although 
we do see an increase in the relative noise with reducing box size, the 
overall noise is much less than the underlying signal. Further, the 
absolute value of the signals are in close agreement as well, the 
largest difference being $\sim15\%$ at $\ell=2000$ and the qualitative 
shape of the power spectra agree well at least for $\ell<7000$. For 
$\ell=10000$ we see a difference of 50\% in the signals, and we are not 
confident of convergence in the interval $7000<\ell < 10000$ (see 
Sections~\ref{conv2}, \ref{conv} for an extensive discussion of 
convergence issues).

\subsection{Impact of spatial resolution}

\begin{figure}
\epsscale{.80}
\plotone{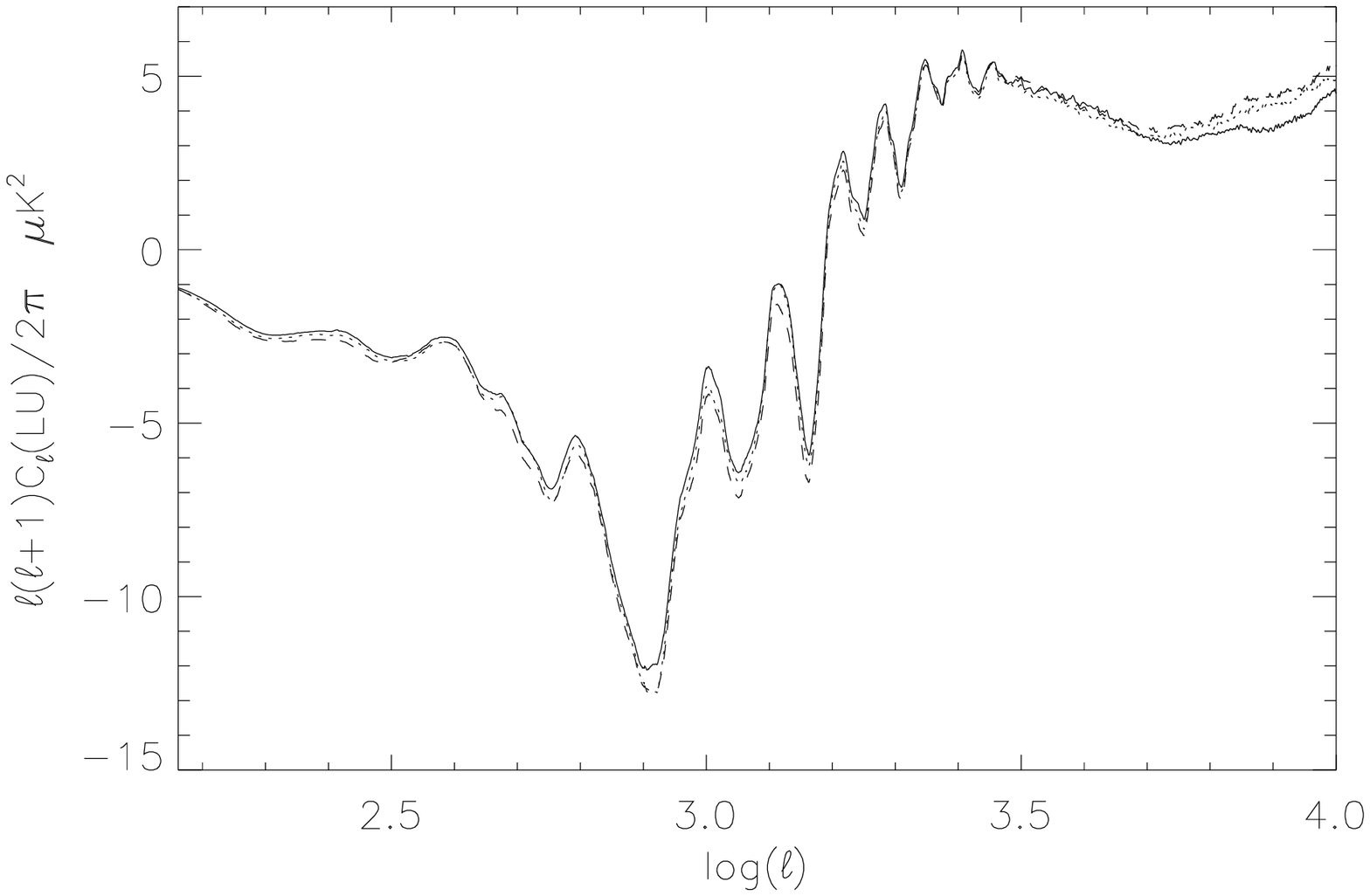}
\caption{LU angular power spectra extracted for RLS\_BA (solid line), 
and 
from other two LSs 
with the same parameters and initial conditions, but differing 
softenings and photon steps where, $\Delta_{ps} = 50 h^{-1} \ 
kpc$ 
and $S_{p} = 24 h^{-1} \ kpc$ (LS\_$\Delta$S1AA, dotted line),
and $\Delta_{ps} = 75 h^{-1} \ kpc$ and $S_{p} = 36 
h^{-1} \ kpc$
(LS\_$\Delta$S2AA, dashed line).
\label{fig6}}
\end{figure}

Having examined sampling issues related to box sizes and preferred 
directions we now address the impact of force softening and the step 
size in the ray-tracing. As a first investigation, we varied parameters 
$S_{p} $ and $\Delta_{ps} $, in lock-step with the remaining RLS 
parameters held fixed. A ratio $E_{res}/\Delta_{ps} \simeq 2.5$, 
identical to that of the RLS simulations has been assumed in all the 
cases, thus maintaining a constant ratio between the photon step and the 
effective resolution (\ie $2.5$ photon steps per effective resolution 
interval). We considered the following new pairs of $S_{p} $ and 
$\Delta_{ps} $: (i) $S_{p} = 24 h^{-1} \ kpc$ and $\Delta_{ps} = 50 
h^{-1} \ kpc$ and, (ii) $S_{p} = 36 h^{-1} \ kpc$ and $\Delta_{ps} = 75 
h^{-1} \ kpc$, and call the LSs associated with these parameters 
LS\_$\Delta$S1AA and LS\_$\Delta$S2AA respectively.  The $C_{\ell}(LU)$ 
for these LSs, along with RLS\_BA for comparison, are given in 
Fig.~\ref{fig6}. Up to $\ell \sim 7000$ the three spectra are very 
similar, although for $\ell > 7000$ we see a small separation in the 
results. For $\ell > 5000$, the trend with decreasing softening and 
photon-step appears to be systematic, a smaller softening leading to 
smaller $C_{\ell}(LU)$ coefficients. Nevertheless, a smaller softening 
does not necessarily mean a more realistic simulation and, consequently, 
we cannot decide which of the three lines of Fig.~\ref{fig6} is closer 
to the true spectrum. Regardless of this issue, any of these lines gives 
a good estimate of the required spectrum for $\ell < 7000 $.

\begin{figure}
\epsscale{.80}
\plotone{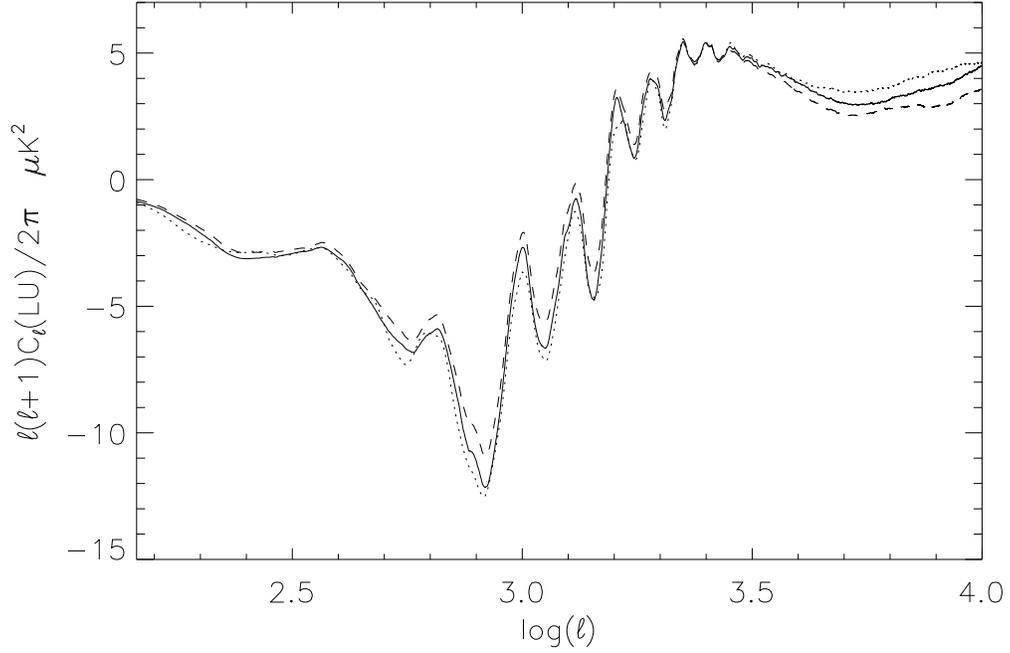}
\caption{Average LU angular power spectra extracted from RLSs (solid 
line), and from other two LSs with the same parameters, excepting the 
photon step $\Delta_{ps}$, whose values are $\Delta_{ps} = 40 h^{-1} \ 
kpc $ 
(dotted line) and $\Delta_{ps} = 12 h^{-1} \ kpc $ (dashed line). The 
dashed, solid, and dotted lines corresponds to the smallest, medium 
(RLS), and greatest $\Delta_{ps}$ values, respectively. Averages have 
been performed by using four appropriate simulations (see text).
\label{fig7}}
\end{figure}

To address the role of the photon step alone, we next held fixed all the 
RLS parameters except the photon step, $\Delta_{ps}$. Along with our 
original choice of $\Delta_{ps}=25 h^{-1} \ kpc$, two 
new values were considered: $\Delta_{ps} = 12 h^{-1} \ 
kpc$ ($\sim E_{res}/5$, corresponding to LSs 
LS\_$\Delta$1AA, LS\_$\Delta$1AB, LS\_$\Delta$1BA, LS\_$\Delta$1BB), 
and 
$\Delta_{ps} = 40 
h^{-1} \ 
kpc$ (a little smaller than $\sim E_{res}$, corresponding to LSs 
LS\_$\Delta$2AA, LS\_$\Delta$2AB, LS\_$\Delta$2BA, LS\_$\Delta$2BB). 
For each of the $\Delta_{ps}$ 
two distinct 
preferred directions, namely $D512A$ and $D512B$, were considered 
along with two distinct sets of initial modes, giving 
rise to four LSs.
The four spectra for each $\Delta_{ps}$ value were 
then averaged and the resulting spectra plotted in Fig. \ref{fig7}. 
Note, that since the error in the lensing deviation integrals is reduced 
with decreasing $\Delta_{ps}$, we can reasonably argue that our best 
estimate of the lensing signal is given by the minimum $\Delta_{ps}$. We 
see that in all three cases the $C_{\ell}(LU)$ quantities are in close 
agreement up to $\ell \sim 7000$. In the $\ell$-interval 
[7,000--10,000], the separations between the three curves increase, with 
a systematic trend of a lower signal with decreasing photon-step. Since 
all these spectra have the same simulation softening it is reasonable to 
conclude that the trend we observed when comparing different $S_p$ and 
$\Delta_{ps}$ may actually be a function of the change in $\Delta_{ps}$. 
Given that the smallest $\Delta_{ps}$ exhibits the lowest signal, we 
thus conclude that our best estimate of the signal actually has the 
smallest correlations for $\ell >7000 $ values. Nonetheless, for $\ell < 
7000$, the RLS closely follows our best estimate, hence for $\ell < 
7000$ we are very confident that the RLS leads to a robust estimate of 
the lensing effect we are studying. Even if the $\Delta_{ps} = 12  
h^{-1} \ kpc$ value leads to the most accurate spectrum 
(which is not certain), the computational cost of constructing these LSs 
is significantly greater than that of the RLSs, while only providing a 
mild improvement in accuracy for $\ell > 7000$.

\begin{figure}
\epsscale{.80}
\plotone{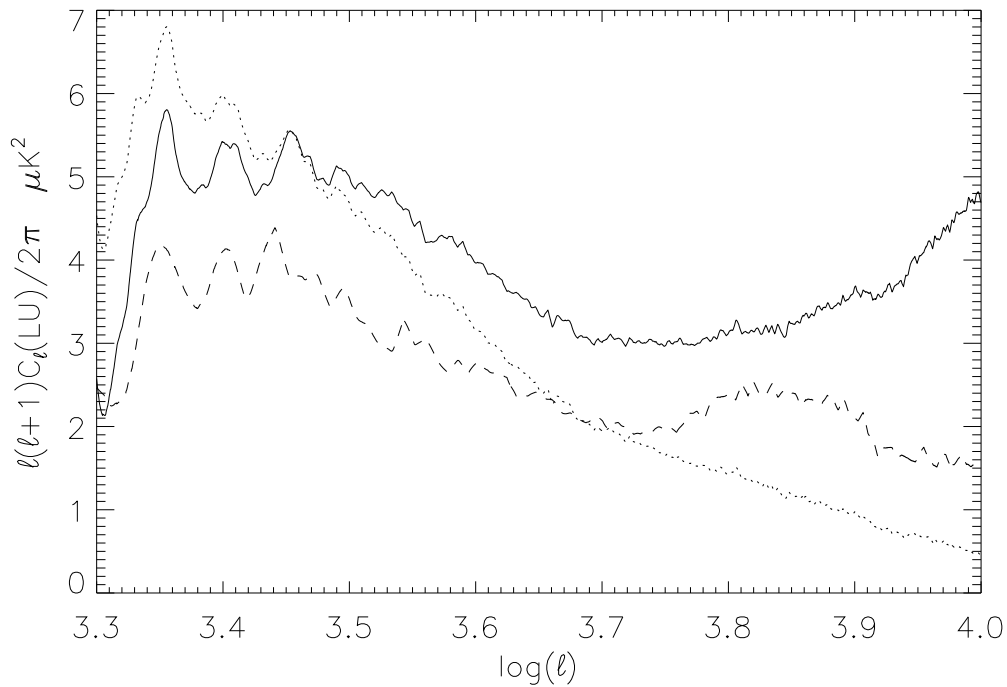}

\caption{LU angular power spectra for the RLS\_AA simulation (solid 
line) as compared to the same simulation but where deflections are 
calculated by including an average over the 8 nearest geodesics (dotted 
line). This reduces the resolution of geodesic method, but maintains the 
same resolution in the gravitational solver. Also shown (dashed line) is 
the highest resolution (but small volume) LS\_HEA simulation. As shown 
in more detail in figure 11, higher resolution reduces the impact of 
discreteness effects, which in turn reduces the asymptotic high $\ell$ 
signal. However, sample variance issues in the LS\_HEA simulation 
preclude us for drawing firm conclusions at present.}
\label{fig10_new}
\end{figure}

We have also attempted to reproduce results similar to that found 
elsewhere using interpolation methods. We ran another simulation using 
RLS\_AA parameters, except that this time the deflections were 
calculated using the 8 nearest neighbours. This keeps the mass 
resolution of the simulation fixed and does not alter the 
gravitational 
calculation, while at the same time lowering the effective resolution 
of 
the lensing methodology. The resulting power spectrum is shown in Fig. 
\ref{fig10_new} and shows a decaying signal at high $\ell$ which is 
similar to that found in earlier work (\eg \cite{db08}). This shows that 
as we degrade the resolution of ray-tracing method we do indeed recover 
previous results. 

\subsection{Impact of mass resolution on convergence}
\label{conv2}
\begin{figure}
\epsscale{.80}
\plotone{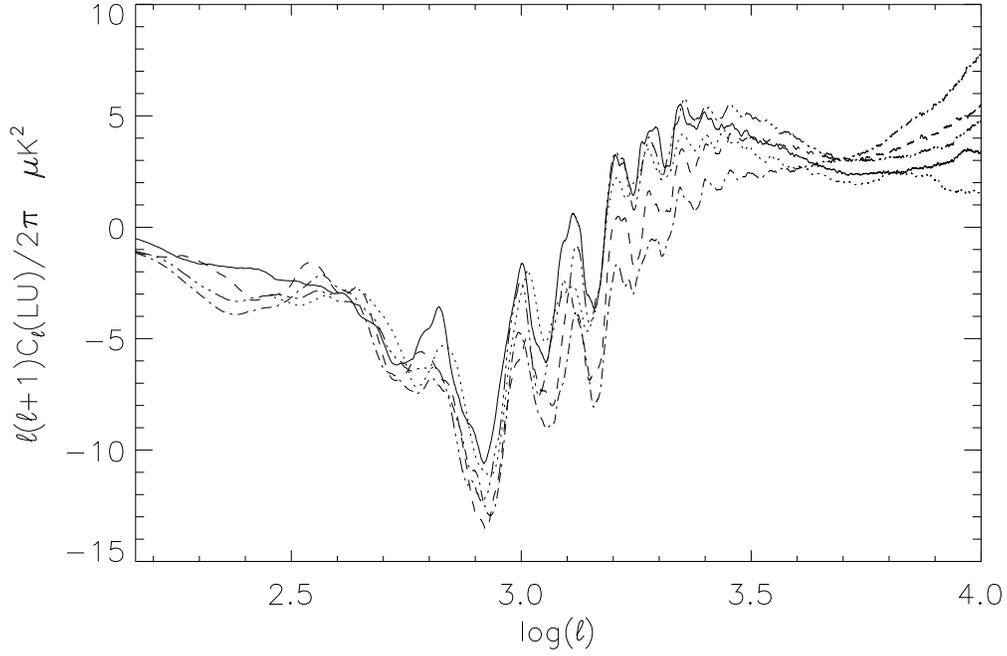}
\caption{Comparison of convergence in the LU angular power spectra for 
different mass resolutions. The dot-dashed line corresponds to the 
$128^3$ simulation LS\_LHA, the dashed line to the $256^3$ simulation 
LS\_MGA, the triple-dot-dashed line to RLS\_AA, the solid line to 
LS\_$\Delta$1AB, and the dotted line to LS\_HEA. 
The simulation boxes are $512h^{-1}$ Mpc except for 
the LS\_HEA simulation which uses 1/8th the volume. The two $512^3$ 
simulations have the same mass resolution but they bracket a range of 
photon-step values (see Table~\ref{tab2} for 
simulation details), a full estimate of the uncertainty in 
the signal at this mass resolution is given in Figure 13. 
\label{fig8new}}
\end{figure}

\begin{figure}
\epsscale{.80}
\plotone{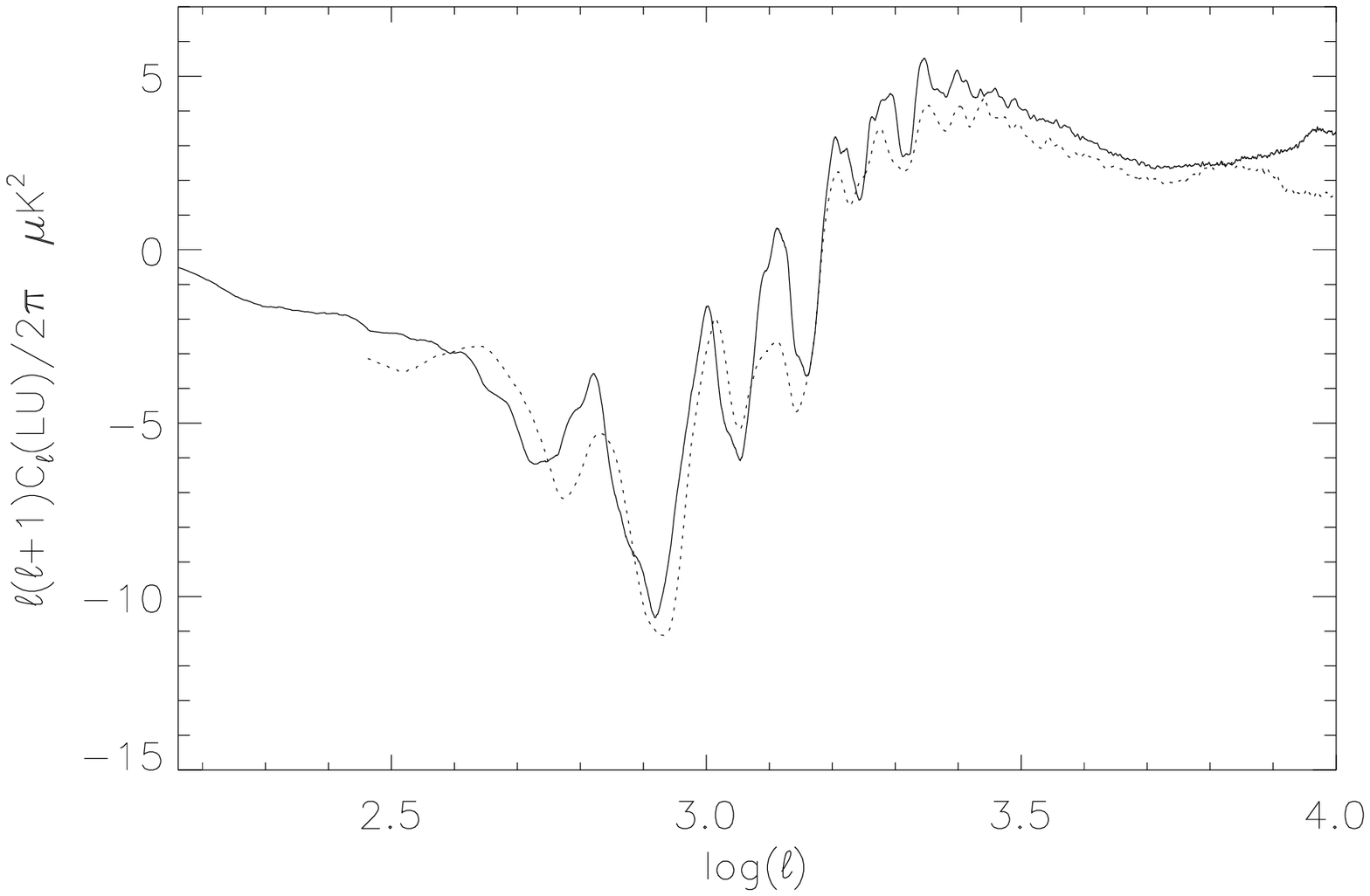}
\caption{LU angular power spectra extracted from
two LS obtained from fully independent initial conditions, but differing
in short-scale resolution.
The dotted line corresponds to LS\_HEA with parameters
$L_{box}=256 h^{-1} \ Mpc$, $N_{p} = 512^{3}$, $N_{c} = 1024$,
$N_{dir}=512$, $S_{p} = 6 \times 10^{-3} h^{-1} \ Mpc$, and
$\Delta_{ps} = 15  h^{-1} \ kpc$. The solid line corresponds to
LS\_$\Delta$1AB which has $\Delta_{ps} = 12 \ kpc$ and the remaining 
parameters
identical to those of the RLSs.
\label{fig8}}
\end{figure}

Since our ray-tracing method samples the local gravitational field on 
scales down to a few tens of kpc, it is important to investigate the 
role of mass resolution on our simulation. While we have investigated 
box size earlier, which at a fixed resolution leads to a change in mass 
resolution, in this section we focus on changing the mass resolution in 
a fixed volume, and also include one additional simulation in a smaller 
volume at our maximum mass resolution.

We ran additional simulations with $128^3$ and $256^3$ particles in the 
RLS box size, namely $512h^{-1}$ Mpc. These simulations are labeled 
LS\_LHA and LS\_MGA in Table~\ref{tab2}. Notable other parameters 
were 
$S_p=48h^{-1}(24h^{-1})$ kpc, $\Delta_{ps}=120h^{-1}(60h^{-1})$ kpc and 
particle masses were $7.0\times10^{12}\, (8.8\times10^{11}) \, M_\odot$   
respectively.  
In Fig.~\ref{fig8new} we plot the 
power spectrum results for these simulations, along with an RLS, another 
$512^3$ simulation with a shorter photon step (LS\_1AB), and one 
final simulation labeled LS\_HEA that uses a volume 1/8th the size 
of an RLS but with the same number of particles. The parameters of 
this LS were $L_{box}=256 h^{-1}$  Mpc, 
$S_{p}=6h^{-1}$  kpc and $\Delta_{ps} = 15 h^{-1}$  kpc.
Note the two $512^3$ simulations have the same mass resolution but they
bracket a range of photon-step values, a full estimate of the 
uncertainty in
the signal at this mass resolution is given in Figure 13, where the 
spectra
of these two cases appear inside the band of uncertainty. 
Due to the different box size of the LS\_HEA run we could not 
maintain the same modes across all runs, and hence we used different 
initial modes at different resolutions. 
Taken together, 
these simulations
span a ratio in mass resolution of 512. 

The plot shows that when PP 
corrections are included there is a resolution-dependent 
up-turn in the power spectrum for $\ell>5000$. Improving the mass 
resolution reduces the 
impact of N-body discreteness on the lensing (\eg \citealt{jai00}) to 
the 
point where 
the
highest resolution simulation (LS\_HEA)
appears to be free of this contamination for
$\ell<10^4$. Turning to lower $\ell$ values, 
the peak of the power spectrum around $\ell\sim 2000$ appears to be 
converged, modulo the variation due to the initial modes, for 
resolutions 
of $512^3$ and greater. The results for $\ell<1000$ also appear to show 
errors consistent with the variation in initial modes observed in 
Section~\ref{modes}, indicating results are converged in this range for 
all 
resolutions. 

\subsection{Is convergence at $\ell > 7000$ possible?}
\label{conv}
To address the question of the correct form of the spectrum at $\ell > 
7000$ we utilize the RLSs and the LS\_HEA simulation.
The 
effective resolution $E_{res} \simeq 30h^{-1} \ kpc$ of the LS\_HEA 
simulation
is half of that attained in the RLSs and thus small scales should be 
represented more accurately. As a result of the reduced 
box size, the angular size of the lensed maps is 
$\Phi_{map}=2.48^{\circ}$ and the angular resolution limit $\Delta_{ang} 
= 
0.29^{\prime} $ ($\ell \sim 37000$) is half that of the RLSs (\ie higher 
resolution).

The resulting spectrum is plotted in Fig.~\ref{fig8} as a dotted line, 
and compared to our previous best estimate for the signal at $\ell > 
7000$, namely that from LS\_$\Delta$1AB that was constructed with 
$\Delta_{ps} = 
12 h^{-1} \ kpc$ and 
the remaining parameters identical to those of the RLSs (the 
preferred direction is D512B). As expected, given that the new spectrum 
is obtained from smaller $2.48^{\circ} \times 2.48^{\circ}$ maps, some 
uncertainties in the $C_{\ell}(LU)$ quantities are observed for $\ell < 
2000$. Note that for the $4.96^{\circ} \times 4.96^{\circ}$ maps used in 
the RLS we expect uncertainties for $\ell < 1000$. However, for $\ell > 
2000 $, the spectrum of the smaller box (dotted line) is expected to be 
more accurate than our previous LS. Although we again observe a 
systematic trend of better resolution producing a lower signal at high 
$\ell > 7000$, the results are nonetheless in good quantitative 
agreement for $2000 < \ell < 7000$. We can hence be confident that our 
RLSs give also a rough but useful estimate of the angular power spectrum 
for $2000 < \ell < 7000$ (supporting our assertion in the previous 
section). Finally, for $7000 < \ell < 10,000$, the smaller box produces 
a decrease, whereas the previous LS increases slightly. This 
unfortunately suggests that the N-body simulations and the CMB lensed 
maps used to derive the RLSs are not good enough for $\ell > 7000$. We 
hence conclude that future simulations with higher resolutions are 
clearly necessary to go beyond $\ell = 7000$.

\section{Discussion and observational implications}
\label{obs}
\begin{figure} 
\epsscale{.80} 
\plotone{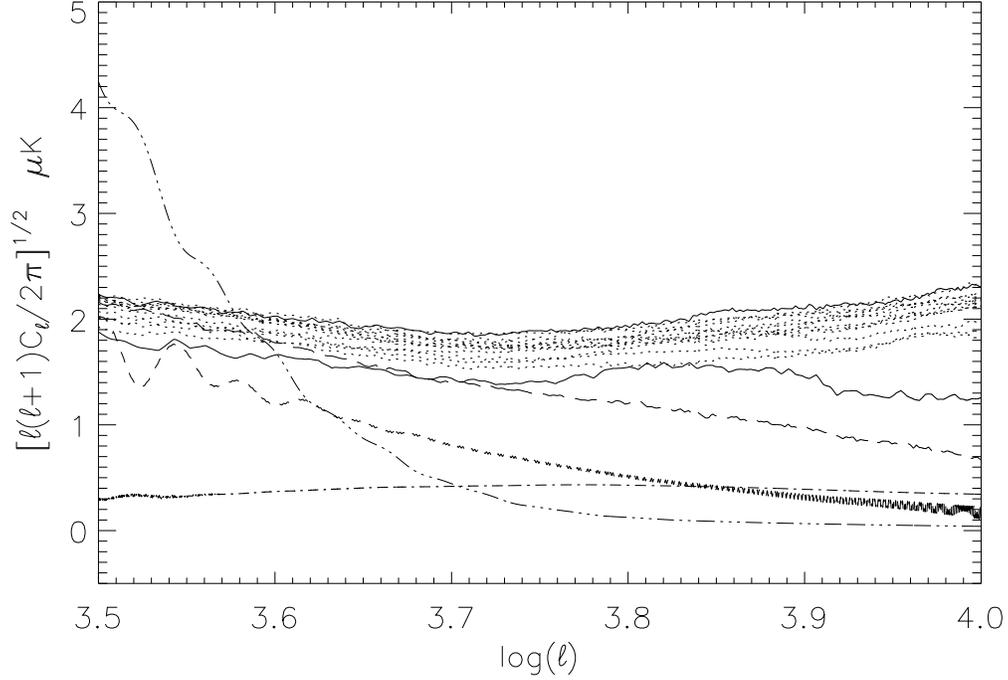} 
\caption{Angular power spectra, in $\mu$K, as functions of $\log(\ell)$. 
Solid and dotted lines represent all the spectra displayed in 
Figs.~\ref{fig3}-\ref{fig8}. The lower solid line displays the same 
spectrum as the dotted line of Fig.~\ref{fig8}, which corresponds to the 
LS\_HEA LS, which is the highest resolution simulation considered. The 
upper solid line shows the same spectrum as the dashed line of 
Fig.~\ref{fig6}, which corresponds to the LS\_$\Delta$S2AA LS, which has 
the same mass resolution as the RLSs but an $S_{p}$ value 3 times 
larger. The long dashed line corresponds to the geodesically averaged 
signal (mimicing reduced resolution), as shown in Figure 10. The 
dash-three-dot line is the CMB spectrum in the absence of lensing, and 
the dash-dot (dashes) line is the AWL (BWL) effect obtained with the 
CMBFAST code in the case of nonlinear lensing including structures with 
sizes $L > 30h^{-1} \ Kpc$. \label{fig9}}
\end{figure}

To bring together all that we have learned thus far, in Fig.~\ref{fig9} 
we plot all the $LU$ angular power spectra from our simulations (excepting
simulation LS\_LHA with $N_{p}=128^{3} $ particles) in the 
$\log(\ell)$-interval (3.5, 4.0), which we reemphasize corresponds to the 
AWL effect described earlier (scales smaller than $42 h^{-1} \ Mpc$ at $z < 6$).
The $AWL$ and the complementary lensing effect 
$BWL$ estimated with CMBFAST are also included in the same plot. Since 
we demonstrated the $CWL$ component is negligible in the interval under 
consideration
(see the bottom left panel of Fig~\ref{fig0}), it is
not represented. The 
lower 
solid line displays the same spectrum as the dotted line of 
Fig.~\ref{fig8}, which corresponds to the LS\_HEA LS, which has the 
smallest value of $S_{p} $ and the highest mass resolution considered 
(eight times better than that of the RLSs). The upper solid line shows 
the same spectrum as the dashed line of Fig.~\ref{fig6}, which 
corresponds to the LS\_$\Delta$S2AA LS, which has the same mass 
resolution as the RLSs but an $S_{p}$ value 3 times larger (and 6 times 
larger than that used in the LS\_HEA LS). Despite the large differences 
in the numerical parameters defining the N-body simulations, we see that 
from $\ell \simeq 3000$ to $\ell \simeq 7000 $, all the spectra lie in a 
narrow band (bounded by the solid lines of Fig.~\ref{fig9}), having a 
width close to half a micro-Kelvin. 
In addition to the simulation lines 
(solid and dotted) we have plotted estimates of the lensing signal based 
upon semi-analytical approaches implemented in the CMBFAST code \citep{lew06}.
By using the nonlinear version of this code with 
a minimum scale of $30h^{-1} \ kpc$ (as in our LS, leading to the lower solid 
line of Fig.~\ref{fig9}), we have estimated the AWL (dash-dots line in 
Fig.~\ref{fig9} and top left panel of Fig.~\ref{fig0}) and 
the BWL components (dashed line in 
Fig.~\ref{fig9} and top right panel of Fig.~\ref{fig0}) of the weak lensing. Finally, 
the CMBFAST code has been also used to calculate 
the CMB angular spectrum, in the absence 
of lensing, for the model described in 
Section~\ref{maps} (dash-three-dots line).

The complementary effect 
$BWL$, not calculated within the simulation, is 
due to scales greater than $L_{max} = 42 h^{-1} \ Mpc$. While the absence of this 
power has significant 
consequences for small $\ell$ values we will now show (with CMBFAST)
that in the interval we are 
examining it is small. 
If the scale $L_{max} = 42 h^{-1} \ Mpc$ 
is placed on the last scattering surface at ($z \simeq 1100 $), it 
subtends an angle of $14^{\prime}.63$. However, in Fig.~\ref{fig9} we are plotting 
$\ell$ values 
corresponding to angles of $\sim 3^{\prime}.4$ and $\sim 1^{\prime}$, \ie 
considerably smaller than the scale set by $L_{max}$ 
on the last scattering surface. 
By this reason, the power of the dashed line, which is
due to the linear scales greater than $L_{max} $, is 
expected to be small
in the $\log(\ell)$-interval of
Fig.~\ref{fig9}. CMBFAST estimates have confirmed this smallness
(dashed line of Fig.~\ref{fig9}). Nevertheless,
the $BWL$ effect is very important for much smaller $\ell$ values
(see bottom right panel of Fig.~\ref{fig0}). 

The complementary $CWL$ effect (see the bottom left panel of  
Fig.~\ref{fig0}) is too small to be significant in the 
$\log(\ell)$-interval (3.5, 4.0). The $AWL$ effect obtained with
CMBFAST (nonlinear version) is shown in the dot-dashed line of
Fig.~\ref{fig9}. This effect appears to be smaller than 
the $AWL$ effect numerically 
estimated here (with AP3M simulations), 
whose power is found to be close to $2 \ \mu K$ for 
$3000 < \ell < 7000$. The total lensing would be obtained
by adding this power and that of the dashed line ($BWL$ 
component). We thus see that the total lensing 
effect has a power of a few $\mu \ K$. 

To compare this result with other simulation work, we turn to the work 
by \cite{db08}.
Since our map-making method does not allow us to plot the lensed and
unlensed
$C_\ell$ we have converted their results (specifically those in their
Figure 4) to the $[\ell (\ell+1)C_\ell(LU)/(2\pi)]^{1/2}$ values we use 
in our paper . In 
performing this
conversion we have multiplied their power spectra by a factor of
$2.90\times10^{13}$ $\mu$K${}^2$ which
includes the effective fractional sky
coverage, $f^{eff}_{sky}$.
We hence
find values of 1.1 $\mu$K at $\ell=4000$, 0.7
$\mu$K at $\ell=5000$, 0.6 $\mu$K at $\ell=6000$ and 0.4 $\mu$K at
$\ell=7000$. This
compares with our own signal of $2.0\pm0.4$ $\mu$K 
across the
entire range. However, we
caution
against
over-interpreting this comparison. Our cosmology and assumptions about
reionization and backgrounds are  slightly different from those in
\cite{db08}, however
it is clear that we find a higher signal at $\ell>1000$ when short range
corrections are included.

As it is apparent from Fig.~\ref{fig9}, the $AWL$ effect clearly 
dominates
the 
primary anisotropies
for $\ell > 4200 $ (\ie the continuous and 
dotted lines are well above the dash-three-dots one). However, the 
Sunyaev-Zel'dovich (SZ) effect is also important at these scales, and 
foregrounds
must be properly accounted for. 
The Berkeley-Illinois-Maryland Association (BIMA),
has reported \citep{daw06} a temperature power 
($[\ell(\ell+1)C_{\ell}/2\pi]^{1/2}$)
of $14.88^{+4.09}_{-4.88} \ \mu K$ at 68 \% confidence, for 
$\ell = 5237 $.
In order to explain $14.88 \ \mu K $ as an SZ effect, one must have 
$\sigma_{8} \simeq
1.03$ \citep{coo02,daw06}, but this is in comparatively strong disagreement with data 
from WMAP (see Section~\ref{sec1}) that lead 
to $\sigma_{8}\simeq
0.82$. Hence, taking into account that the SZ temperature power
scales as $\sigma_{8}^{3.5} $ \citep{kom02,bon05,daw06,sie09}, 
the SZ effect could 
only explain around the 44.5 \% of the most likely BIMA value ($14.88 \ 
\mu K$).
The Cosmic Background Imager (CBI) observations, \citet{bon05}, 
cover the multipole range $400 < \ell < 4000$ and, in a recent analysis
\citep{sie09}, it is argued that, if the CBI excess power observed at 
very small angular scales is explained as a SZ effect for a certain
$\sigma_{8}$ value, then a power close to $10 \ \mu K$, marginally 
compatible with 
BIMA measurements at $\ell = 5237 $, may be explained with the
same value of $\sigma_{8}$.

The level of power we predict for the weak lensing ($AWL$ plus $BWL$ components) 
may thus be a 
missing link, simultaneously ensuring the compatibility of WMAP, CBI and 
BIMA observations. We have predicted a lensing effect of a few 
micro-Kelvin for these large angular scales, and hence a smaller SZ 
component (a smaller value of $\sigma_{8}$) could lead to simultaneous 
compatibility with WMAP, CBI, and BIMA observations. Aside from the 
large errors bars on the BIMA measurement, we are also cautious about 
making firm claims since lensing and SZ effects are essentially produced 
by the same structure distributions (galaxy clusters and sub-structures 
involving dark matter and baryons) and, consequently, these effects must 
be strongly correlated. This implies that the spectra of these two 
effects must be superposed in an unknown way (not merely added). In 
practice, this superposition might be analyzed in detail with 
ray-tracing through hydrodynamical simulations (including both baryons 
and dark matter). More work along these lines is clearly necessary.

\section{Conclusions}

\label{discuss}

Our first conclusion, in agreement with other work, is that PM 
simulations are inefficient for calculating CMB lensing due to strongly 
nonlinear structures (see Fig.~\ref{fig2} and comments in 
Section~\ref{comparison}). While it is not impossible to simulate these 
effects with a PM code the required resolution is so large that the 
additional short-scale resolution provided by AP3M (or other high 
resolution N-body technique) is far more efficient at capturing the 
lensing effect. Hence our development of a combined parallel 
AP3M-ray-tracing code is a necessary step to estimate the lensing signal 
at high $\ell$.

Only the $AWL$ effect (see Section~\ref{nbody}) produced by scales 
smaller than $42h^{-1}  \ Mpc$ between redshifts $z_{in} = 6 $ and $z_{end}=0$
has been estimated with AP3M simulations.
This choice is appropriate for the following reasons: (1) all the strongly 
nonlinear scales are taken into account; (2) 
omitting scales greater than $42h^{-1}  \ Mpc$ 
makes our ray-tracing procedure efficient (see Section~\ref{nbody}); (3) 
the effect produced by scales greater than $42h^{-1}  \ Mpc $ 
($BWL$ in Section~\ref{nbody}) can be studied by using
the linear version of CMBFAST; and, (4) the lensing due to scales smaller
than $42h^{-1}  \ Mpc$, at redshift $z>6 $ ($CWL$ in 
Section~\ref{nbody}), 
can be computed using standard semi-analytical
methods implemented in CMBFAST.

Given that our study has strongly focused on the $AWL$ signal, we
exhaustively investigated
the numerical issues involved when estimating this signal. Our 
main conclusions can be summarized as follows: 
\begin{itemize}
\item the lensing contribution between $z=0.2$ and $z=0$
is negligible
\item for each RLS, a few realizations 
suffice to get a good average
$C_{\ell}(LU)$ spectrum and, moreover, each single simulation gives a very good spectrum 
rather similar to the average one
\item RLSs, which are essentially identical 
except for the preferred directions, give similar spectra
(the ray-tracing procedure has little variance)
\item simulation boxes of $1024h^{-1} \ Mpc $ are not fully necessary for $\ell > 
1000$; 
however, for $\ell < 1000$, these large sizes 
lead to the most reliable spectra and,
moreover, such sizes should lead to very good spectra 
in the $\ell$-interval                                
($1000$--$2000$)

\item simulations in boxes of $512h^{-1} \ Mpc $ 
lead to good $C_{\ell}(LU)$ spectra for $1000 < \ell < 7000 $

\item for $2000 < \ell < 7000 $, all the RLSs lie in a region of width 
$\sim$0.5 $\mu$K, indicating that the RLSs give consistent estimates of 
the 
signal in this range 

\item the signal in the range $4000 < \ell < 7000 $ is $2.0\pm0.4 \; 
\mu$K, which is $\sim \! 1.4$ $\mu$K higher than that found elsewhere 
(\eg \citealt{db08})

\end{itemize}

Despite some simulation uncertainties, our code and technique have lead 
to a robust estimate of the lensing effect in the $\ell$-interval 
$(4200,7000)$, where it clearly dominates the primary anisotropy. 
Moreover, the estimated power is larger than that obtained with 
semi-analytical methods and the CMBFAST code. We thus suggest that the 
resulting value of a few micro-Kelvin may explain the excess power at 
high $\ell$ in the BIMA and CBI observations. This conclusion is 
supported by recent studies based on the Millennium simulation 
\citep{car08}, where the authors have reported a small contribution from 
nonlinearity at $\ell \simeq 4100 $. However, the methods of 
\citet{car08} have been designed to build all-sky lensed maps, and do 
not have the resolution necessary to perform an accurate estimate of the 
weak lensing by strongly nonlinear structures in the $\ell$-interval 
where we have found our main effect.

Our direct estimation of the potential gradients at photon positions 
using PP corrections from the local dark matter particles appears to be 
the main origin of the difference between our results and other research 
relying either on planes or grid interpolations. However, we emphasize 
that differences only occur at the large $\ell>2000$ values we have been 
investigating. Our method employs extremely fine time resolution, namely 
that used by AP3M simulations and also a very good angular resolution 
(see above). Due to current limitations in our method it is not possible 
for us to determine the precise role of temporal resolution in an 
accurate lensing calculation. 

It is well known that on small scales baryons do not follow the dark
matter distribution. Thus, while we have attempted to be as accurate as
possible in this dark matter simulation, we are now probing scales where
contributions from baryons are beginning to become significant. An
investigation of the impact of baryons by \cite{ji06} considered two
types of simulations, the first with non-radiative baryons while the
second included dissipation and star formation. They found that for
$1000<l<10,000$ an effect of between 1\% to 10\% on the weak-lensing
shear angular power spectrum,
as calculated from dark matter alone, was possible. The largest
difference was produced in the run with dissipation and star formation,
where a 10\% increase in the (shear) $C_\ell$s at $\ell=10000$ was
observed.
Future work is definitely necessary to determine the impact of this
physics on lensing statistics. We plan to conduct simulations with
baryons and feedback processes both to identify its impact on the signal
we find, and also to systematically evaluate the combined impact of the
SZ effect and weak lensing. 

For calculations that are accurate to high $\ell$ it seems to be 
necessary to move CMB photons through the simulation box while 
structures are evolving, which ensures the spatial gradients are 
accurately calculated on the photon positions (test particles). Once the 
N-body code has been modified to compute spatial gradients at particle 
test positions, there are no theoretical or technical reasons to take a 
temporal resolution different from that defined by the simulation time 
step. From the theoretical point of view, the time step resolution is 
obviously most compatible with the N-body technique (and thus takes into 
account the entire evolution of the simulation). From the technical 
point of view, the use of the time step resolution requires the minimum 
memory cost as the number of test particles between two successive 
times, although large, is minimized. Because of these considerations, 
temporal resolution is not a parameter to be varied in our calculations.

Our AP3M code adapted to CMB lensing calculations can be run 
for different values of the parameters defining the LSs; hence, 
this code allows us to
see how the resulting angular power spectra depend on the
parameters defining both the N-body simulation and the ray-tracing 
procedure.

\acknowledgments

We thank the anonymous referee for suggestions that improved the clarity 
and content of the paper. M.J. 
Fullana, J.V. Arnau and D. S\'aez 
acknowledge the financial support of the Spanish
Ministerio de Educaci\'on y Ciencia, MEC-FEDER project
FIS2006-06062. RJT and HMPC acknowledge funding by individual
Discovery Grants from NSERC. RJT is also supported by grants from
the Canada Foundation for Innovation and the Canada Research Chairs
Program. HMPC acknowledges the support of the Canadian Institute for
Advanced Research. Simulations were performed at the {\em Computational
Astrophysics Laboratory} at Saint Mary's University.


\end{document}